\documentstyle[12pt,amsfonts]{article}
\def\one{1\hskip-.37em 1}  
\def\l{\lambda}
\def\D{{\cal D}}
\def\O{{\cal O}}
\def\a{\alpha}

\def\K{{\cal K}}
\def\Hi{{\frak H}}
\def\H{{\cal H}}
\def\E{{\rm E}\hskip-.55em{\rm I}\,}

\def\ir{{\rm I}\hskip-.2em{\rm R}}
\def\half{\textstyle{\frac{1}{2}}}
\def\quarter{\textstyle{\frac{1}{4}}}

\def\ra{\rightarrow}
\def\tint{{\textstyle{\int}}}
\def\T{{\sf T}}
\def\d{\partial}
\def\o{\overline}
\def\e{\epsilon}       
\def\bn{\begin{eqnarray}}     
\def\en{\end{eqnarray}}       
\def\<{\langle}
\def\>{\rangle}

\def\{{\lbrace}
\def\}{\rbrace}
\title{Quantization of Constrained Systems\footnote{To appear in
 the proceedings of the 39th  Schladming Winter School on ``Methods of
Quantization'', February 26-March 4, 2000, Schladming, Austria.}} 
\author{John R. Klauder\\
Departments of Physics and Mathematics\\
University of Florida\\
Gainesville, FL  32611}
\date{}  
\begin{document}
\maketitle
\begin{abstract}
The present article is primarily a review of 
the projection-operator approach to quantize systems 
with constraints. We
study the quantization of systems with general first- and second-class
 constraints from the point of view of coherent-state,
phase-space path integration, and show that all such cases may be treated,
 within the original classical phase space, by using suitable path-integral
 measures for the Lagrange multipliers which ensure that the quantum system
 satisfies the appropriate quantum constraint conditions.  Unlike
 conventional methods, our procedures involve no $\delta$-functionals of the
 classical constraints, no need for dynamical gauge fixing of first-class
 constraints nor any average thereover, no need to eliminate second-class
 constraints, no potentially ambiguous determinants, as well as no need to
 add auxiliary dynamical variables expanding the phase space beyond its
 original classical formulation, including no ghosts. Besides several 
pedagogical examples, we also study: (i)
the quantization procedure 
for reparameterization invariant models, (ii) systems for which the 
original set of Lagrange mutipliers are elevated to the status of dynamical 
variables and used to define an extended dynamical system which is completed 
with the addition of suitable conjugates and new sets of constraints and their
associated Lagrange multipliers, (iii) special examples of alternative but 
equivalent formulations of given 
first-class constraints, as well as (iv) a comparison of both regular and 
irregular constraints.
\end{abstract}


\section{INTRODUCTION}
\subsection{Initial comments}
 The quantization of systems with constraints is important conceptually as 
well as
 practically.  Principal techniques for the quantization of such
 systems involve conventional operator techniques \cite{dir}, path integral
 techniques in terms of the original phase space variables \cite{fad},
 extended operator techniques involving ghost variables in addition to the
 original variables and extended path integral techniques also including
 ghost fields (see, e.g., \cite{dym,gov,hen}). However, these standard 
approaches are generally 
not unambiguous and may exhibit certain difficulties in application. A 
recent review
\cite{shapr} carefully analyzes these traditional methods and details their 
weaknesses
as well as their strengths. 

Canonical quantization generally requires the use of Cartesian coordinates 
and not more general
coordinates \cite{dir2}. Therefore, whenever we 
consider a dynamical system without any
 constraints whatsoever, we assume that the phase space of the unconstrained
 system is flat  and admits a standard quantization of its canonical
 variables either in an operator form or in an equivalent path integral
 form. Next, suppose constraints exist, which, for the sake of
 discussion, we choose as a closed set of first-class constraints;
 extensions to treat more general constraints are presented in later sections.
 Whenever there are constraints the original set of variables is no longer
 composed solely of physical variables but now contains some unphysical
 variables as well. While such variables cause little concern from a classical
 standpoint, they are viewed as highly unwelcome from a quantum standpoint 
inasmuch
 as one generally wants to quantize only physical variables. Thus it is 
often deemed
 necessary to eliminate the unphysical variables leaving only the true
 physical degrees of freedom. Quantization of the true degrees of freedom is
 supposed to proceed as in the initial step. In the general case, however,
 a quantization of the remaining degrees of freedom is not straightforward 
or perhaps not even  possible
 because the physical (reduced) phase space is non-Euclidean  
meaning that an
  obstruction has arisen where none existed before. An 
 obstruction generally precludes the existence of self-adjoint (observable!) 
canonical
 operators satisfying the canonical commutation relations. In path integral
 treatments, such obstructions arise from the introduction of delta
 functionals that enforce the classical constraints and the concomitant need
 to introduce subsidiary delta functionals to select a compatible dynamical
 gauge in order to introduce a canonical sympletic structure on the physical  
phase space that generally is not flat.  
These are fundamental problems that seem
 difficult to overcome.

This article reviews a middle ground in the quantization procedure of systems
with constraints which may be called the {\it projection-operator, 
coherent-state approach}.
Briefly stated, quantization of the original, unconstrained variables 
proceeds without
obstruction or ambiguity, while constraints are enforced by means of a 
well-chosen
projection operator projecting the original Hilbert space onto the physical 
Hilbert
subspace. This conservative framework is presented in the form of a 
phase-space
path integral with the help of coherent states (which, while convenient, 
are not necessary).
The difference between the present approach and other functional integral 
methods
may be attributed to an alternative choice for the integration measure 
for the
Lagrange multiplier variables. The present approach may be traced from 
\cite{kkk}. In 
addition, some aspects of the projection operator approach have been 
presented in unpublished work
of Shabanov \cite{sha71}; see also \cite{sha}.

\subsection{Classical backround}
For our initial discussion, let us briefly review the classical theory of 
constraints.
Let $\{p_j,q^j\}$, $1\leq j\leq J$, denote a set of dynamical variables, 
$\{\l^a\}$, $1\leq a\leq A$, a set of Lagrange multipliers, and 
$\{\phi_a(p,q)\}$ a set of constraints. Then the dynamics of a 
constrained system may be summarized in the form of an action principle 
by means of the classical action (summation implied)
  \bn I=\tint[p_j{\dot q}^j-H(p,q)-\l^a\phi_a(p,q)]\,dt\;.  
\label{e1}\en
 The resultant equations that arise from the action read
  \bn  &&{\dot q}^j=\frac{\d H(p,q)}{\d p_j}+\l^a\frac{\d\phi_a(p,q)}{\d
p_j}\equiv\{q^j,H\}+\l^a\{q^j,\phi_a\}\;,\nonumber\\
  &&{\dot p}_j=-\frac{\d H(p,q)}{\d q^j}-\l^a
\frac{\d\phi_a(p,q)}{\d q^j}\equiv\{p_j,H\}+\l^a\{p_j,\phi_a\}\;,\nonumber\\
  &&\phi_a(p,q)=0\;,  \label{e2}\en
where $\{\cdot,\cdot\}$ denotes the Poisson bracket. The set of 
conditions $\{\phi_a(p,q)=0\}$ defines the {\it constraint hypersurface}. 
If the constraints satisfy 
\bn &&\{\phi_a(p,q),\phi_b(p,q)\}=c_{ab}^{\;\;\;\;c}\,
\phi_c(p,q) \;,
\label{e3}\\
  &&\{\phi_a(p,q),H(p,q)\}=h_a^{\;\;b}\,\phi_b(p,q)
\label{e4}\;,  \en
then we are dealing with a system of first-class constraints. If the 
coefficients $c_{ab}^{\;\;\;\;c}$ and $h_a^{\;\;b}$ are constants, then 
it is a closed system of first-class constraints; if they are suitable 
functions of the variables $p,q$, then it is called an open first-class 
constraint system. If (\ref{e3}) fails, or (\ref{e3}) and (\ref{e4}) fail, 
then the constraints are said to be second class (see below). 

For first-class constraints it is sufficient to impose the constraints at 
the initial time inasmuch as the equations of motion will ensure that the 
constraints are fulfilled at all future times. Such an initial imposition 
of the constraints is called an {\it initial value equation}. Furthermore, 
the Lagrange multipliers are not determined by the equations of motion; 
rather the solutions of the equations of motion depend on them. By specifying 
the Lagrange multipliers, the solution can be forced to satisfy an 
additional (``gauge'') condition. Observable quantities are gauge 
invariant and, hence, do not depend on the gauge abritrariness. For 
second-class constraints, on the other hand, the Lagrange multipliers 
are determined by the equations of motion in such a way that the 
constraints are satisfied for all time. 

In the remainder of this section we review standard quantization 
procedures for systems with closed first-class constraints, both of the 
operator and path integral variety, pointing out some problems in each 
approach. 

\subsection{Quantization first: Standard \\operator quantization}
For a system of closed first-class constraints we assume (with $\hbar=1$) 
that 
\bn &&[\Phi_a(P,Q),\Phi_b(P,Q)]=ic_{ab}^{\;\;\;\;c}\,\Phi_c(P,Q)\;,\\
  &&[\Phi_a(P,Q),{\cal H}(P,Q)]=ih_a^{\;\;b}\,\Phi_b(P,Q)\;,  \en
where $\Phi_a$ and $\H$ denote self-adjoint constraint and Hamiltonian 
operators, respectively. Following Dirac \cite{dir}, we adopt the 
quantization prescription given by
  \bn i{\dot W}(P,Q)=[W(P,Q),{\cal H}(P,Q)]  \en
where $W$ denotes a general function of the kinematical operators $\{Q^j\}$ 
and $\{P_j\}$ which are taken as a self-adjoint, irreducible representation 
of the commutation rules $[Q^j,P_k]=i\delta^j_k\one$, with all other 
commutators vanishing. The equations of motion hold for all time $t$, 
say $0<t<T$. On the other hand, the conditions
  \bn  \Phi_a(P,Q)|\psi\>_{ phys}=0  \en
to select the physical Hilbert space are imposed only at time $t=0$ as 
the analog of the initial value equation; the quantum equations of motion 
ensure that the constraint conditions are fulfilled for all time.

The procedure of Dirac has potential difficulties if zero lies in the 
continuous spectrum of the constraint operators for in that case there are 
no normalizable solutions of the constraint condition. We face the same 
problem, of course, and our resolution is discussed below.
\subsection{Reduction first: Standard \\path integral quantization}
Faddeev \cite{fad} has given a path integral formulation in the case of 
closed first-class constraint systems as follows. The formal path integral
 \bn &&\int\exp\{i\tint_0^T[p_j{\dot q}^j-H(p,q)-\l^a\phi_a(p,q)]\,dt\}\,
\D p\,\D q\,\D\l \nonumber\\
  &&\hskip1.5cm=\int\exp\{i\tint_0^T[p_j{\dot q}^j-H(p,q)]\,dt\}\,
\delta\{\phi(p,q)\}\,\D p\,\D q \label{e9}\en
may well encounter divergences in the remaining integrals. Therefore, 
subsidiary conditions in the form $\chi^a(p,q)=0$, $1\leq a\leq A$, are 
imposed picking out (ideally) one gauge equivalent point per gauge orbit, 
and in addition a  factor in the form of the Faddeev-Popov determinant is 
introduced to formally preserve canonical covariance. The result is the path 
integral 
 \bn  \int\exp\{i\tint_0^T[p_j{\dot q}^j-H(p,q)]\,dt\}\,\delta\{\chi(p,q)\} 
\det(\{\chi^a,\phi_b\})\delta\{\phi(p,q)\}\,\D p\,\D q\,.  \en
This result may also be expressed as
 \bn \int\exp\{i\tint_0^T[p^*_j{\dot q}^{*j}-H^*(p^*,q^*)]\,dt\}\,\D p^*\,
\D q^*\;,  \label{g11}\en
namely, as a path integral
over a reduced phase space in which the $\delta$-functionals have been used 
to eliminate $2A$ integration variables. 

The final expression generally involves an integral over a non-Euclidean 
phase space for which the conventional definition of the path integral is 
typically ill defined.  Thus this widely used prescription is not without 
its difficulties.

\subsection{Quantization first $\not\equiv$ reduction first}
The two schemes illustrated in the preceding sections are different in 
principle. In the initial case, one quantizes {\it first} and reduces 
{\it second;} in the latter case, one reduces {\it first} and quantizes 
{\it second}. For certain systems the results of these different procedures 
are the same, but that is not universally the case, as we now proceed to 
illustrate.

Let us consider the example of a single degree of freedom specified by the 
classical action
  \bn  I=\tint[p{\dot q}-\l(p^2+q^4-E)]\,dt\;.  \en
Observe that the classical Hamiltonian vanishes and there is a single 
constraint. The question we pose is: For what values of $E$, $E>0$, is 
the quantum theory nontrivial?

On the one hand, according to the procedure of Dirac, the physical Hilbert 
space is either empty or 
one-dimensional, spanned by the nonvanishing eigenvector $|\psi_n\>$ that 
satisfies
   \bn   (P^2+Q^4)|\psi_n\>=E_n\,|\psi_n\>\;,   \en
for $E_n$ one of the purely discrete eigenvalues for the ``Hamiltonian'' 
$P^2+Q^4$. 

On the other hand, the procedure of Faddeev leads initially to 
  \bn  \int e^{i\tint p\,dq}\,\delta\{p^2+q^4-E\}\,\D p\,\D q\;.  \en
Next, we fix a gauge, e.g., $p=0$, in which case the reduced phase space 
propagator is given by
  \bn && \int e^{i\tint p\,dq}\,\delta\{p^2+q^4-E\}\,\Pi(4q^3)\,
\delta\{p\}\,\D p\,\D q\nonumber\\
&&\hskip1cm =0\;,  \en
which vanishes due to cancellation between the term with $q>0$ and the 
term with $q<0$. Note that the symbol $\Pi$ denotes a formal multiplication 
over all time points. An alternative evaluation
may be given if we allow only the term with $q>0$, which is achieved by 
instead using 
  \bn&& \int e^{i\tint p\,dq}\,\delta\{p^2+q^4-E\}\,\delta\{p\}\,\D p\,
\D q^4  \nonumber\\
  &&\hskip1cm =\int\delta\{q^4-E\}\,\D q^4 \nonumber\\
  &&\hskip1cm =1\;.   \en
Either of these choices imposes {\it no} restriction on $E$ whatsoever. 
Ignoring the nonphysical
nature of the variables involved, one might possibly impose the condition
    \bn  \oint p\,dq=2\pi n \;,  \en
leading to a Bohr-Sommerfeld spectrum, which for this problem is incorrect.
(The reader is encouraged to examine alternative choices of gauge.)

{\bf Remark:} It is instructive in this example to note that the 
Faddeev-Popov determinant $\Delta=\Pi(4q^3)$ 
and the reduced phase space is the single point $(p,q)=(0,E^{1/4})$. The 
point $(p,q)=(0,-E^{1/4})$ corresponds 
to a Gribov copy. 

Clearly, in this case, reduction before quantization has led to the wrong 
result. 
Some workers may assert that such errors are merely
``order $\hbar$ corrections''. Although true, this argument cannot be used 
to defend 
the general procedure since the role of a
quantization procedure, after all, should be to determine the {\it correct} 
spectrum 
for a specific problem, not a spectrum that
is potentially incorrect even in its {\it leading order}. Examples of other 
work which arrive at the same conclusion are given in \cite{z1}.

\subsection{Outline of the remaining sections}
In the following section, Sec.~2, we present an overview of the 
{\it projection operator approach} to constrained system
quantization with an emphasis on coherent-state representations. Section 3 
deals with coherent-state path integrals without gauge fixing
for closed first-class constrained systems. Extensions to general 
constraints such as open first-class or second-class systems
are the subject of Sec.~4. Section 5 is devoted to selected examples of 
first-class systems, while Sec.~6 concentrates on two rather special 
applications. Finally, in Sec.~7 we comment on some other applications of 
the projection operator approach that have not been discussed in this paper.
\section{OVERVIEW OF THE PROJECTION \\OPERATOR APPROACH TO CON-\\STRAINED 
SYSTEM QUANTIZATION}
\subsection{Coherent states}
Canonical quantization is consistent only for Cartesian phase space
 coordinates \cite{dir2}, and we assume that our original and unconstrained
 set of classical dynamical variables fulfill that condition. Then, for each
 classical coordinate $q^j$ and momentum $p_j$, $1\leq j\leq J$, we may
 introduce associated self-adjoint canonical operators $Q^j$ and $P_j$, 
acting in a separable Hilbert space $\frak H$, and which
 satisfy, in units where $\hbar=1$, the canonical commutation relations
 $[Q^j,P_k]=i\delta^j_k\,\one$, with all other commutation relations
 vanishing. With the fiducial vector $|0\>\in \Hi$ a suitable normalized 
state---typically the
 ground state of a (unit-frequency) harmonic oscillator (but not
 always!)---we introduce the {\it canonical coherent states} (see, e.g., 
\cite{klasud,klska}) 
  \bn |p,q\>\equiv e^{-iq^jP_j}\,e^{ip_jQ^j}\,|0\> 
\label{e17}\;, \en
for all $(p,q)\in\ir^{2J}$, where $p=\{p_j\}$ and $q=\{q^j\}$. These
 states admit a resolution of unity in the form \cite{mckl}
  \bn \one=\int\,|p,q\>\<p,q|\,d\mu(p,q)\;,\hskip1cm
    d\mu(p,q)\equiv d^J\!p\,d^J\!q/(2\pi)^J\;,  \en
integrated over $\ir^{2J}$.

The unit operator resides in the Hilbert space $\Hi$ of the unconstrained
 system. We may conveniently represent this Hilbert space as follows. We
 first introduce the {\it reproducing kernel} $\<p'',q''|p',q'\>$ as the 
overlap
 matrix element between any two coherent states. This expression is a
 bounded, continuous function that characterizes a (reproducing kernel
 Hilbert space) representation of $\Hi$ appropriate to the unconstrained
 system as follows. A dense set of vectors in the associated functional
 Hilbert space is given by vectors of the form 
 \bn\psi(p,q)\equiv\<p,q|\psi\>=\sum_{l=1}^L\,\alpha_l\,\<p,q|p_{(l)},q_
{(l}\>\;,\en
for arbitrary sets $\{\alpha_l\}$ and $\{p_{(l)},q_{(l)}\}$ with $L<\infty$.
 The inner product of two such vectors is given by
\bn&&(\psi,\xi)\equiv\sum_{l,m=1}^{L,M}\,\alpha^*_l\,\beta_m\,\<p_{(l)},q
_{(l)}|{\o p}_{(m)},{\o q}_{(m)}\>\label{e20}\\
&&\hskip1cm=\int\psi(p,q)^*\xi(p,q)\,d\mu(p,q)\;,  \label{e21}\en
where $\xi$ is a second function defined in a manner analogous to $\psi$.
A general vector in the functional Hilbert space is defined by a Cauchy
 sequence of such vectors, and all such vectors are given by bounded,
 continuous functions. The first form of the inner product applies in
 general only to vectors in the dense set, while the second form of the
 inner product holds for arbitrary vectors in the Hilbert space.
We shall have more to say below regarding reproducing kernels and reproducing 
kernel Hilbert spaces.

\subsection{Constraints}
Now suppose we introduce constraints into the quantum theory \cite{kkk}. 
In particular,
 we assume that $\E$ denotes a {\it projection operator onto the constraint
 subspace}, i.e., the subspace on which the quantum constraints are
 satisfied (in a sense to be defined below), and  which is called the 
physical Hilbert space $\Hi_{phys}\equiv\E\,\Hi$.
Later we shall discuss examples of $\E$. 
 Hence, if $|\psi\>\in\Hi$ denotes a general vector in the original
 (unconstrained) Hilbert space, the vector $\E|\psi\>\in\Hi_{phys}$
 represents its component within the physical subspace. As a Hilbert space,
 the physical subspace also admits a functional representation by means of a
 reproducing kernel which may be taken as $\<p'',q''|\E|p',q'\>$. In the
 same manner as before, it follows that a dense set of vectors in
 $\Hi_{phys}$ is given by functions of the form
\bn  \psi(p,q)\equiv\<p,q|\E|\psi\>=\sum_{l=1}^L\,\alpha_l\,\<p,q|\E|p_{(l)
},q_{(l)}\>\;,  \en
for arbitrary sets $\{\alpha_l\}$ and $\{p_{(l)},q_{(l)}\}$ with $L<\infty$.
 The inner product of two such vectors is given by
\bn  &&(\psi,\xi)\equiv\sum_{l,m=1}^{L,M}\,\alpha^*_l\,\beta_m\,\<p_{(l)},q
_{(l)}|\E|{\o p}_{(m)},{\o q}_{(m)}\>\nonumber\\
&&\hskip1.2cm=\int\psi(p,q)^*\xi(p,q)\,d\mu(p,q)\;. \label{e23} \en
Again, a general vector in the functional Hilbert space is defined by means
 of a Cauchy sequence, and all such vectors are given by bounded, continuous
 functions. Note well, in the case illustrated, that even though
 $\E\,\Hi\subset\Hi$, the functional representation of the unconstrained and
 the constrained Hilbert spaces are {\it identical}, namely by functions of
 $(p,q)\in\ir^{2J}$, and the form of the inner product is {\it identical} in
 the two cases. This situation holds even if $\Hi_{phys}$ is one
 dimensional!

The relation between the self-adjoint constraint operators $\Phi_a$, 
$1\le a\le A$,
 $A<\infty$, and the projection operator $\E$ may take several different 
forms. Unless
otherwise specified, we shall assume that $\Sigma_a\Phi_a^2$ is self adjoint 
and 
that
  \bn \E=\E(\Sigma_a\Phi_a^2\le\delta(\hbar)^2)\;, \en
where $\delta=\delta(\hbar)$ ({\it not} a Dirac $\delta$-function!) is a 
regularization parameter which is chosen 
in accord with rules to be discussed below.

\subsection{Dynamics for first-class systems}
Suppose further that the Hamiltonian $\H$ respects the first-class character
 of the constraints. It follows in this case that $[\E,\H]=0$ or stated
 otherwise that
  \bn  e^{-i\H t}\E\equiv\E \,e^{-i\H t}\E\equiv\E \,e^{-i(\E\,\H\,\E\,)\, 
t}\E\;. 
\label{e26}  \en
 Dynamics in the physical
 subspace is then fully determined by the propagator on $\Hi_{phys}$, which
 is  given in the relevant functional representation by
  \bn  \<p'',q''|e^{-i\H t}\E|p',q'\>\;.  \label{e27} \en
In (\ref{e27}) we have achieved a fully gauge invariant propagator without 
having to
 reduce the range or even the number of the original classical variables nor
 change the original form of the inner product on the functional Hilbert
 space representation. Any observable $\O$---$\H$ included---satifies 
$[\E,\O]=0$, and relations
 similar to (\ref{e26}) follow with $\H$ replaced by $\O$.

\subsection{Zero in the continuous spectrum}
The foregoing scenario has assumed that the appropriate $\Hi_{phys}$ is
 given by means of a projection operator $\E$ acting on the original Hilbert
 space. This situation holds true whenever the set of quantum constraints
 admits zero as a common point in their discrete spectrum; in that case $\E$
 defines the subspace where the constraints all vanish. That situation may
 not always hold true, but even in case zero lies in the continuous spectrum
 for some or all of the constraints, a suitable result may generally be
 given by matrix elements of a sequence of rescaled projection operators,
 say $c_\delta\,\E$, $c_\delta>0$, as $\delta\ra0$. Specifically, we
 consider the limit of a sequence of reproducing kernels
 $c_\delta\,\<p'',q''|\E|p',q'\>$, which---if the limit is a nonvanishing
 continuous function---defines a new reproducing kernel, and thereby a new
 reproducing kernel Hilbert space, within which the appropriate constraints
 are fulfilled. In such a limit certain variables may cease to be relevant
 and as a consequence the local integral representation of the inner 
product, if any,
 may require modification. On the other hand, the definition of the inner
 product by sums involving the reproducing kernel will always hold. We
refer to the result of such a limiting operation as a {\it reduction of
the reproducing kernel}.
A simple example should help clarify what we mean by a reduction of the 
reproducing kernel.

Consider the example
 \bn &&\hskip-.4cm\<p'',q''|\E|p',q'\>\nonumber\\
&&\hskip.4cm=\pi^{-1/2}\int_{-\delta}^\delta\exp[-\half(k-p'')^2+ik(q''-q')
- -\half(k-p')^2]\,dk\;,  \label{f27}\en
where $\E=\E(P^2\le\delta^2)$, which defines a reproducing kernel for any 
$\delta>0$ that corresponds 
to an infinite dimensional Hilbert space.  (If $\delta=\infty$ the
 result is the usual canonical coherent state overlap and characterizes the
 unconstrained Hilbert space.) 
If we take the limit of the expression as it stands 
as $\delta\ra0$, the result will vanish. What we need to do is extract the 
{\it germ} of the 
projection operator as we let $\delta$ go to zero. Therefore, let us first 
multiply this 
expression by $\pi^{1/2}/(2\delta)$ [$c_\delta$ in this case] and take the 
limit $\delta\ra0$.
The result is the expression
 \bn \K(p'';p')=e^{-\half(p''^2+p'^2)}\;,  \label{f28}\en
which has become a reproducing kernel that characterizes a 
{\it one}-dimensional Hilbert space with every functional representative 
proportional to $\chi_o(p)\equiv\exp(-p^2/2)$. This one-dimensional Hilbert 
space representation also admits a local integral representation for the 
inner product given by
\bn  (\chi,\chi)=\tint |\chi(p)|^2\,dp/\sqrt{\pi}\;.  \en

In the present case, it is clear that one may reduce the reproducing kernel 
even further
by choosing $p=c$, an arbitrary but fixed constant. This kind of 
reduction---in which the latter
reproducing kernel Hilbert space is equivalent to the former reproducing 
kernel Hilbert
space---is analogous to choosing a gauge in the classical theory. We shall 
see another
example of this latter kind of reduction later.

The example presently under discussion is also an important one inasmuch as 
it illustrates how a constraint operator with its zero lying in the continuous
spectrum is dealt with in the coherent-state, projection-operator approach. 
Some 
other approaches to deal
 with the problem of zero in the continuous spectrum 
may be traced from \cite{zero}.

\subsection{Alternative view of continuous zeros}
If $\delta\ll1$ in (\ref{f27}), then it may be approximately evaluated as
  \bn  &&\<p'',q''|\E|p',q'\>\nonumber\\
    &&\hskip1cm=\pi^{-1/2}\delta\,e^{-\half(p''^2+p'^2)}\,\frac{\sin[
\delta(q''-q')]}{\delta(q''-q')}+O(\delta^2) 
\label{g}\;.\en
When $\delta=10^{-1000}$, or some other extremely tiny factor, it is clear 
that for all 
practical purposes it is sufficient to accept
just the first term in (\ref{g}), ignoring the term $O(\delta^2)$, as the 
``reduced'' reproducing kernel.
The resultant expression is indeed a proper reproducing kernel for which 
inner products are given with the
full set of integration variables and the normal integration range. So long 
as $q$ values are ``normal
sized'', e.g., $|q|<10^{500}$ in the present case, there is no practical 
distinction  between the space of functions
generated by (\ref{f28}) and that generated by (\ref{g}). In other words, 
if $\delta$ is chosen extremely close to zero,
but still positive, it is not actually necessary to take the limit 
$\delta\ra0$ in order to do practical calculations.
Even though this is the case, we shall for the most part in the examples we 
study take a full reduction by first rescaling the reproducing 
kernel (by an appropriate factor $c_\delta$) and then taking the limit 
$\delta\ra0$.

\section{COHERENT STATE PATH INTEGRALS\\WITHOUT GAUGE FIXING}
As introduced above,  canonical coherent states may be defined by the relation
 \bn  |p,q\>\equiv e^{-iq^jP_j}\,e^{ip_jQ^j}\,|0\>\;,  \label{e30}\en
for all $(p,q)$, where the fiducial vector $|0\>$ traditionally denotes a 
normalized, unit frequency, harmonic oscillator ground state, and the 
coherent states admit a resolution of unity in the form
 \bn  \one=\tint\,|p,q\>\<p,q|\,d\mu(p,q)\;,\hskip1.5cm d\mu(p,q)\equiv 
d^J\!p\,d^J\!q/(2\pi)^J\;,  \en
where the integration is over $\ir^{2J}$. Note that the integration domain 
and the form of the 
measure are unique. 

Based on such coherent states, we introduce the upper symbol for a general 
operator ${\cal H}(P,Q)$, 
   \bn  H(p,q)\equiv\<p,q|{\cal H}(P,Q)|p,q\>=\<p,q|:H(P,Q):|p,q\> \;  \en
which is related to the normal-ordered form as shown. (N.B. Some workers 
would call $H(p,q)$ the lower symbol.) If ${\cal H}(P,Q)$ denotes the 
quantum Hamiltonian, then we shall adopt $H(p,q)$ as the classical 
Hamiltonian. We also note that an important one-form generated by the 
coherent states is given by
$i\<p,q|d|p,q\>=p_j\,dq^j$. 

Using these quantities, and the time ordering operator $\sf T$, the coherent 
state path integral for the  propagator generated by the time-dependent 
Hamiltonian ${\cal H}(P,Q)+\l^a(t)\Phi_a(P,Q)$ is readily given by
\bn \label{e40} &&\hskip-.4cm\<p'',q''|{\sf T}e^{-i\tint_0^T[{\cal H}(P,Q)
+\l^a(t)\Phi_a(P,Q)]\,dt}|p',q'\>\nonumber\\
&&\hskip.2cm=\lim_{\epsilon\ra0}\int\prod_{l=0}^N\<p_{l+1},q_{l+1}|
e^{-i\epsilon({\cal H}+\l^a_l\Phi_a)}\,|p_l,q_l\>\prod_{l=1}^N\,
d\mu(p_l,q_l)\nonumber\\
&&\hskip.2cm=
 \int\exp\{i\tint[i\<p,q|(d/dt)|p,q\>-\<p,q|{\cal H}+\l^a(t)\Phi_a|p,q\>]
\,dt\}\,\D\mu(p,q)\nonumber\\
&&\hskip.2cm= {\cal M}\int\exp\{i\tint[p_j{\dot q}^j-H(p,q)-\l^a(t)
\phi_a(p,q)]\,dt\}\,\D p\,\D q \;.  \en
Here, in the second line, we have set $\epsilon\equiv T/(N+1)$, made 
a Trotter-product like approximation to the evolution operator, repeatedly 
inserted the resolution of unity, and set $p_{N+1},q_{N+1}=p'',q''$ and 
$p_0,q_0=p',q'$. In the third and fourth lines we have formally 
interchanged the continuum limit and the integrations, and written for 
the integrand the form it would assume for continuous and differentiable 
paths ($\cal M$ denotes a formal normalization constant). The result 
evidently depends on the chosen form of the functions $\{\l^a(t)\}$.
 
\subsection{Enforcing the quantum constraints}
Let us next introduce the quantum analog of the initial value equation. 
For simplicity we assume that the constraint operators form a compact 
group; more general situations are dealt with below. In that case 
  \bn \label{e41}\E\,\equiv\tint e^{-i\xi^a\Phi_a(P,Q)}\,\delta\xi=
\E(\Phi_a=0, \;\;1\le a\le A)=\E(\Sigma_a\Phi^2_a=0) \en
defines a {\it projection operator} onto the subspace for which $\Phi_a=0$ 
provided that $\delta\xi$ denotes the normalized, $\tint\delta\xi=1$, group 
invariant measure. It follows from (\ref{e41}) that
  \bn  \hskip1.5cm e^{-i\tau^a\Phi_a}\E\,=\E\,\;.  \en
We now project the propagator (\ref{e40}) onto the quantum constraint subspace
which leads to the following set of relations
 \bn &&\hskip-1cm\int\<p'',q''|{\sf T}e^{-i\tint[{\cal H}+\l^a(t)\Phi_a]
\,dt}\,|{\o p}',{\o q}'\>\<{\o p}',{\o q}'|\E\,|p',q'\>\,d\mu({\o p}',
{\o q}')\nonumber\\
&&=\<p'',q''|{\sf T}e^{-i\tint[{\cal H}+\l^a(t)\Phi_a]\,dt}\,\E\,|p',q'\>
\nonumber\\
&&=\lim\,\<p'',q''|[\prod^{\leftarrow}_l(e^{-i\epsilon{\cal H}}
e^{-i\epsilon\l^a_l\Phi_a})]\,\E\,|p',q'\>\nonumber\\
&&=\<p'',q''|e^{-iT{\cal H}}e^{-i\tau^a\Phi_a}\,\E\,|p',q'\>\nonumber\\
&&=\<p'',q''|e^{-iT{\cal H}}\,\E\,|p',q'\>\;,  \en
where $\tau^a$ incorporates the functions $\l^a$ as well as the structure 
parameters $c_{ab}^{\;\;\;\;c}$ and $h_a^{\;\;b}$.
Alternatively, this expression has the formal path integral representation
 \bn \label{e45}\int\exp\{i\tint[p_j{\dot q}^j-H(p,q)-\l^a(t)\phi_a(p,q)]
\,dt-i\xi^a\phi_a(p',q')\}\,\D\mu(p,q)\,\delta\xi\;.  \en
On comparing (\ref{e40}) and (\ref{e45}), we observe that {\it after 
projection onto the quantum constraint subspace the propagator is entirely 
independent of the choice of the Lagrange multiplier functions. In other 
words, the projected propagator is gauge invariant.} 

We may also express the physical (projected) propagator in a more general 
form, namely,
\bn &&\hskip-1cm\int\exp\{i\tint[p_j{\dot q}^j-H(p,q)-\l^a(t)\phi_a(p,q)]
\,dt\}\,\D\mu(p,q)\,\D C(\l)\nonumber\\
&&\hskip.1cm=\<p'',q''|e^{-iT{\cal H}}\,\E\,|p',q'\>  \en 
provided that $\tint\D C(\l)=1$ and that such an average over the functions 
$\{\l^a(t)\}$ introduces (at least) one factor $\E\,$. 

\subsection{Reproducing kernel Hilbert spaces}
The coherent-state matrix elements of $\E$ define a fundamental kernel
 \bn \K(p'',q'';p',q')\equiv\<p'',q''|\E|p',q'\>\;,  \en
which is a bounded, continuous function
for any projection operator $\E$, especially including the unit operator. 
It follows that $\K(p'',q'';p',q')^*=\K(p',q';p'',q'')$ as well as
  \bn \sum_{k,l=1}^K\alpha^*_k\alpha_l\K(p_k,q_k;p_l,q_l)\geq0\en
for all sets $\{\alpha_k\}$, $\{(p_k,q_k)\}$, and all $K<\infty$. The last 
relation is an automatic consequence of the complex conjugate property and 
the fact that
\bn \K(p'',q'';p',q')=\int \K(p'',q'';p,q)\,\K(p,q;p',q')\,d\mu(p,q)\en
holds in virtue of the coherent state resolution of unity and the properties 
of $\E$. As noted earlier, the function $\K$ is called the {\it reproducing 
kernel} and the Hilbert space it engenders is termed a {\it reproducing 
kernel Hilbert space} \cite{aron}. A dense set of elements in the Hilbert 
space is given by functions of the form
 \bn \psi(p,q)=\sum_{k=1}^K \alpha_k\K(p,q;p_k,q_k)\;,  \en
and the inner product of this function has two equivalent forms given by
\bn &&\hskip-1.2cm (\psi,\psi)=\sum_{k,l=1}^K\alpha^*_k\alpha_l
\K(p_k,q_k;p_l,q_l)\\
&&=\tint\psi(p,q)^*\psi(p,q)\,d\mu(p,q)\;.  \en
The inner product of two distinct functions may be determined by 
polarization of the norm squared \cite{rsz}. Clearly, the entire Hilbert 
space is characterized by the reproducing kernel $\K$. Change the kernel 
$\K$ and one changes the representation of the Hilbert space. Following a 
suitable limit of the kernel $\K$, it is even possible to change the 
{\it dimension} of the Hilbert space, as already illustrated earlier.

\subsection{Reduction of the reproducing kernel}
Suppose the reproducing kernel depends on a number of variables and 
additional parameters. We can generate new reproducing kernels from a 
given kernel by a variety of means. For example, the expressions
 \bn &&\K_1(p'';p')=\K(p'',c;p',c)\;,\\
   &&\K_2(p'';p')=\tint f(q'')^*f(q')\K(p'',q'';p',q')\,dq''\,dq'\;,\\
 &&\K_3(p'',q'';p',q')=\lim\K(p'',q'';p',q')  \en
each generate a new reproducing kernel provided the resultant function 
remains continuous. In general, however, the inner product in the Hilbert 
space generated by the new reproducing kernel is only given by an analog 
of (\ref{e20}) and not by (\ref{e21}), although frequently some sort of 
local integral representation for the inner product may exist.

Let us offer an example of the reduction of a reproducing kernel that is 
a slight generalization of the earlier example. Let the expression
  \bn \label{e56}&&\hskip-.8cm\<p'',q''|\E|p',q'\>\equiv\nonumber\\
&&\hskip-.6cm\pi^{-J/2}\int_{-\delta}^\delta\cdots\int_{-\delta}^\delta\,
\exp[-\half(k-p'')^2+ik\cdot(q''-q')-\half(k-p')^2]\,d^J\!k  \en
denote a reproducing kernel for any $\delta>0$.  In the present case it 
follows that
$\E\equiv\Pi_{j=1}^J\E(-\delta\le P_j\le\delta)$.
When $\delta\ra0$, then (\ref{e56}) vanishes. However, if we first multiply 
by
 $\delta^{-J}$---or more conveniently by $\pi^{J/2}(2\delta)^{-J}$---before
 taking the limit, the result becomes
\bn  \label{e57}\lim_{\delta\ra0}{\pi^{J/2}}{(2\delta)^{-J}}\<p'',q''|
\E|p',q'\>
=\exp(-\half\,p''^{\,2})\exp(-\half\,p'^{\,2})\;,  \en
which is continuous and therefore denotes the reproducing kernel for some
 Hilbert space. Note that the classical variables $q''$ and $q'$ have
 disappeared, which on reference to (\ref{e30}) implies that all 
``$P_j=0$''.  In the present example, the 
resultant Hilbert space is 
 one dimensional, and the inner product may be given either by
a sum as in (\ref{e20}) involving the $p$ variables alone or by a local 
integral
 representation now using the measure $\pi^{-J/2}\,d^J\!p$, namely,
   \bn  (\chi,\chi)=\tint|\chi(p)|^2 \pi^{-J/2}\,d^J\!p\;.  \en
This example
 illustrates the case where the constraints are ``$P_j=0$'', for all $j$, a
 situation where zero lies in the continuous spectrum.

We may also use this example to illustrate how {\it several} constraints 
may be replaced by a {\it single} constraint. The several constraints 
``$P_j=0$'', for all $j$, were first approximated by the regularized 
constraints $P_j^2\le\delta^2$, $\delta>0$, for all $j$.
Alternatively, we may also regularize the constraints in the form 
$\Sigma_jP_j^2\le\delta^2$. Furthermore, if we use $\E=\E(\Sigma_jP_j^2
\le\delta^2)$, then it is clear that a new prefactor, also proportional 
to $\delta^{-J}$, can
be chosen so that (\ref{e57}) again emerges as $\delta\ra0$. 

\subsection{Single regularized constraints}
Clearly, the {\it set} of real classical constraints $\phi_a=0$, $1
\le a\le A$, is equivalent to the {\it single} 
classical constraint $\Sigma_a\phi_a^2=0$. Likewise, the set of (idealized) 
quantum constraints ``$\Phi_a|\psi\>_{phys}=0$'', $1\le a\le A$, where 
each $\Phi_a$ is self adjoint, is equivalent to the single (idealized) 
quantum constraint 
``$\Sigma_a\Phi_a^2|\psi\>_{phys}=0$'', where we further assume that 
$\Sigma_a\Phi_a^2$ is a self-adjoint operator. 
In general, however, the only solution of the idealized quantum 
constraint is the zero vector, $|\psi\>_{phys}=0$.

To overcome this difficulty, we relax the idealized quantum constraint 
and instead generally adopt the regularized form of the constraint given 
by $|\psi\>_{phys}\in\H_{phys}\equiv\E\,\H$, where
 \bn \E=\E(\Sigma_a\Phi_a^2\le\delta(\hbar)^2)\;.  \en
Here $\delta(\hbar)$ is a regularization parameter and the inequality means 
that in a spectral resolution of 
$\Sigma_a\Phi_a^2\equiv\tint_0^\infty\lambda\,dE(\lambda)$ that 
  \bn \E\equiv \int_0^{\delta(\hbar)^2}\!dE(\lambda)=E(\delta(\hbar)^2)\;. \en
Let us examine three basic examples.

First, let zero be in the discrete spectrum of $\Sigma_a\Phi_a^2$. Then, it 
follows that there exists a $\delta_1(\hbar)^2$ such that for all 
$\delta(\hbar)^2$, $0<\delta(\hbar)^2<\delta_1(\hbar)^2$, then 
$\E(\Sigma_a\Phi_a^2\le\delta(\hbar)^2)=\E(\Sigma_a\Phi_a^2=0)$. 

Second, if $\Sigma_a\Phi_a^2$ has its zero in the continuum, then 
$\E(\Sigma_a\Phi_a^2\le\delta^2)$ is infinite dimensional for all 
$\delta>0$, but $\E$ vanishes weakly as $\delta\ra0$. For such cases we 
consider $c_\delta\E$ and choose the sequence $c_\delta$ to weakly extract 
the {\it germ} of $\E$ as $\delta\ra0$, just as in the examples illustrated 
above.

Third, in a case to be studied later, suppose that zero is {\it not} in the 
spectrum of the operator $\Sigma_a\Phi_a^2$. Since $\Sigma_a\phi_a^2=0$ 
classically, it follows that spectral values of $\Sigma_a\Phi_a^2$ are 
$o(\hbar^0)$ close to zero. A relevant example discussed later is where 
$\Phi_1=P$ and $\Phi_2=Q$. Then $\E(P^2+Q^2\le\hbar)=|0\>\<0|$ is a 
one-dimensional projection operator onto the harmonic oscillator ground 
state $|0\>$. Observe in this case that $\delta(\hbar)^2=\hbar$, which 
vanishes when $\hbar\ra0$; note also that we cannot reduce this parameter 
further since $\E(P^2+Q^2<\hbar)\equiv 0$. Thus, in some cases, whether we 
use ``$\le$'' or ``$<$'' in the inequality defining the projection operator 
can make a real difference.

The three types of examples discussed above illustrate three qualitatively 
different behaviors possible for the
projection operator $\E$.
As we proceed, we shall find the use of a single regularized constraint 
will be an important unifying principle in treating
the most general multiple constraint situation imaginable.

\subsection{Basic first-class constraint example}
Consider the system with two degrees of freedom, a vanishing Hamiltonian, 
and a single constraint, characterized by the action
 \bn I=\tint[\half(p_1{\dot q}_1-q_1{\dot p}_1+p_2{\dot q}_2-q_2{\dot p}_2)
- -\l(q_2p_1-p_2q_1)]\,dt\;,  \en
where for notational convenience we have lowered the index on the $q$ 
variables.
Note that we have chosen a different form for the kinematic part of the 
action which amounts to a change of phase for the coherent states, and in 
particular a factor of $e^{ipq/2}$ has been introduced on the right side 
of (\ref{e17}), or, equivalently, both generators appear in the same 
exponent. It follows that
 \bn &&{\cal M}\int\exp\{i\tint[\half(p_1{\dot q}_1-q_1{\dot p}_1+
p_2{\dot q}_2-q_2{\dot p}_2)-\l(q_2p_1-p_2q_1)]\,dt\}\nonumber\\
  &&\hskip3cm\times\,\D p\,\D q\,\D C(\l)\nonumber\\
&&\hskip2cm=\<p'',q''|\E\,|p',q'\>\;,  \en
where we choose
  \bn  \E\,=(2\pi)^{-1}\int_0^{2\pi}e^{-i\xi(Q_2P_1-P_2Q_1)}\,d\xi=
\E\,(L_3=0)\;.  \en
Based on the fact \cite{klasud} that
 \bn \<p'',q''|p',q'\>=\exp(-\half|z''_1|^2-\half|z''_2|^2+z''^*_1z'_1+
z''^*_2z'_2-\half|z'_1|^2-\half|z'_2|^2)\;,  \en
where $z'_1\equiv(q'_1+ip'_1)/\sqrt{2}$, etc., it is straightforward to 
show that
 \bn \label{e65} &&\<p'',q''|\E\,|p',q'\>  =\exp(-\half|z''_1|^2-\half|
z''_2|^2-\half|z'_1|^2-\half|z'_2|^2)\nonumber\\
 &&\hskip3.3cm\times I_0(\,(z''^{*2}_1+z''^{*2}_2)^{1/2}(z'^2_1+
z'^2_2)^{1/2}\,)\;,\en
with $I_0$ a standard Bessel function. We emphasize again that although 
the Hilbert space has been strictly reduced by the introduction of $\E\,$, 
the reproducing kernel (\ref{e65}) leads to a reproducing kernel Hilbert 
space with an inner product having the same number of integration variables 
and domain of integration as in the unconstrained case.

\section{APPLICATION TO GENERAL \\CONSTRAINTS}
\subsection{Classical considerations}
When dealing with a general constraint situation it will typically happen 
that the self-consistency of the equations of motion may determine some or 
all of the Lagrange multipliers in order for the system to remain on the 
classical constraint hypersurface. For example, if the Hamiltonian attempts 
to force points initially lying on the constraint hypersurface to leave that 
hypersurface, then the Lagrange multipliers must supply the necessary forces 
for the system to remain on the constraint hypersurface. 

We may elaborate on this situation as follows. Since $\phi_a(p,q)=0$ for 
all $a$ defines the constraint hypersurface, it is also necessary, for all 
$a$, that 
  \bn {\dot \phi}_a(p,q)\equiv \{\phi_a(p,q),H(p,q)\}+\lambda^b(t)
\{\phi_a(p,q),\phi_b(p,q)\}\equiv0  \label{e58}\en
also holds on the constraint hypersurface.
If the Poisson brackets fulfill the conditions given in (\ref{e3}) and 
(\ref{e4}), then it follows 
that ${\dot\phi}_a(p,q)\equiv 0$ on the constraint hypersurface
for any choice of the Lagrange multipliers $\{\lambda^a(t)\}$. This is the 
case for first-class constraints, and to
obtain specific solutions to the dynamical equations it is necessary to 
specify some choice of the Lagrange multipliers, i.e., to select a gauge. 
However, if (\ref{e3}), or (\ref{e3}) and (\ref{e4}) do {\it not} hold on 
the constraint hypersurface, the situation changes. For example, let us 
first assume that
(\ref{e4}) holds but that 
  \bn \Delta_{ab}(p,q)\equiv \{\phi_a(p,q),\phi_b(p,q)\}  \en
is a {\it non}singular matrix on the constraint hypersurface. In this case 
it follows that we must choose $\lambda^a(t)\equiv0$ for all $a$ to satisfy
(\ref{e58}). More generally,  we must choose 
  \bn  \lambda^a(t)\equiv -(\Delta^{-1}(p,q))^{ab}\,\{\phi_b(p,q),
\,H(p,q)\} \en
in order that (\ref{e58}) will be satisfied. When the Lagrange multipliers 
are not arbitrary but rather must be
specifically chosen in order to keep the system on the constraint 
hypersurface, then we say that we deal with second-class constraints. 
Of course, there are also intermediate situations where part of the 
constraints are first class while some are second class; in this case 
the matrix $\Delta_{ab}(p,q)$ would be singular but would have a nonzero 
rank on the constraint hypersurface. 

{\bf Remark:} It is useful to also imagine solving the differential 
equation (\ref{e58}) as a computer might do it, namely, by an iteration 
procedure. In particular, we could imagine evolving by a small time step 
$\e$ by the first (Hamiltonian) term, then using the second (constraint) 
term to choose $\lambda^a$ at that moment to force the system back onto 
the constraint hypersurface, and afterwards continuing this procedure 
over and over. A proper solution can be obtained this way by taking the 
limit of these approximate solutions as $\e\ra0$. An analogue of this 
procedure will be used in our quantum discussion.

There is also a third situation that may arise, namely constraints that 
are {\it first} class from a classical point of view but are {\it second} 
class quantum mechanically. Such constraints would arise if 
\bn  \Delta_{ab}(p,q)=Y_{ab}^{\;\;\;c}(p,q)\,\phi_c(p,q)\;,  \en
where, for the sake of convenience, we assume that the quantities 
$Y_{ab}^{\;\;\;c}(p,q)$ are all uniformly bounded away from zero and 
infinity, i.e., $0<C\le Y_{ab}^{\;\;\;c}(p,q)$ $\le D<\infty$. In that 
case $\Delta_{ab}(p,q)$ would vanish on the constraint hypersurface 
classically. Quantum mechanically, the expression for the commutator is 
proportional to $\hbar$ and may be taken as
    \bn  i[\Phi_a(P,Q),\,\Phi_b(P,Q)]= \half[Y_{ab}^{\;\;\;c}(P,Q)\,
\Phi_c(P,Q)+\Phi_c(P,Q)\,Y_{ab}^{\;\;\;c}(P,Q)] \;.\en
If we assume that ``$\Phi_a(P,Q)|\psi\>_{phys}=0$'', then self-consistency 
requires that 
  ``$[\Phi_c(P,Q),\,Y_{ab}^{\;\;\;c}(P,Q)]|\psi\>_{phys}=0$'',
an expression which is now proportional to $\hbar^2$. If this expression 
vanishes it causes no problem; if it does not vanish one says that there 
is a  ``factor ordering problem'' or an ``anomaly''. 
As Jackiw has often stressed, it would be preferable to call an anomaly 
``quantum mechanical symmetry breaking'', a phrase which more accurately 
describes what it is and what it does. Whatever it is called, the resultant 
quantum constraints are second class even though they were classically 
first class. As is well known, gravity falls into just this category.

In this section we take up the quantization of these more general 
situations involving both first and 
second class constraints  \cite{kkk}.
\subsection{Quantum considerations}
As in previous sections, we let $\E\,$ denote the projection operator onto 
the quantum constraint subspace. Motivated by the classical comments given 
above we consider the quantity
\bn \label{e66}\lim\,\<p'',q''|\E\,e^{-i\epsilon{\cal H}}\E\,
e^{-i\epsilon{\cal H}}\cdots\E\,e^{-i\epsilon{\cal H}}\E\,|p',q'\> \en
where the limit, as usual, is for $\epsilon\ra0$. The physics behind this 
expression is as follows. Reading from right to left we first impose the 
quantum initial value equation, and then propagate for a small amount of 
time ($\epsilon$). Next we recognize that the system may have left the 
quantum constraint subspace, and so we project it back onto that subspace, 
and so on over and over. In the limit that $\epsilon\ra0$ the system 
remains within the quantum constraint subspace and (\ref{e66}) actually 
leads to
 \bn \label{e67}\<p'',q''|\E\,e^{-iT(\E\;{\cal H}\E\;)}\E\,|p',q'\>\;, \en
which clearly illustrates temporal evolution entirely within the quantum 
constraint subspace. If we assume that $\E\,{\cal H}\E\,$ is a self-adjoint 
operator, then we conclude that (\ref{e67}) describes a unitary time 
evolution within the quantum constraint subspace. 

The expression (\ref{e66}) may be developed in two additional and 
alternative ways. First, we repeatedly insert the resolution of unity 
in such a way that (\ref{e66}) becomes
 \bn \label{e68}\lim\,\int\prod_{l=0}^N\<p_{l+1},q_{l+1}|\E\,
e^{-i\epsilon{\cal H}}\E\,|p_l,q_l\>\prod_{l=1}^Nd\mu(p_l,q_l)\;. \en
We wish to turn this expression into a formal path integral, but the 
procedure used previously relied on the use of unit vectors, and the 
vectors $\E\,|p,q\>$ are generally not unit vectors. Thus, let us rescale 
the factors in the integrand introducing \bn  |p,q\>\!\>\equiv\E\,|p,q\>/\|
\E\,|p,q\>\| 
 \en which are unit vectors. If we let $M''\equiv\|\E\,|p'',q''\>\|$, 
$M'\equiv\|\E\,|p',q'\>\|$, and observe that $\|\E\,|p,q\>\|^2=
\<p,q|\E\,|p,q\>$, it follows that (\ref{e68}) may be rewritten as
 \bn M''M'\lim\,\int\prod_{l=0}^N\<\!\<p_{l+1},q_{l+1}|
e^{-i\epsilon{\cal H}}|p_l,q_l\>\!\>\prod_{l=1}^N\<p_l,q_l|\E\,|p_l,q_l\>
\,d\mu(p_l,q_l)\;. \en
This expression is represented by the formal path integral
\bn M''M'\int\exp\{i\tint[i\<\!\<p,q|(d/dt)|p,q\>\!\>-\<\!\<p,q|
{\cal H}|p,q\>\!\>]\,dt\}\,\D_E\mu(p,q)\;,  \en
where the new formal measure for the path integral is defined in an 
evident fashion from its lattice prescription. We can also reexpress 
this formal path integral in terms of the original bra and ket vectors 
in the form 
 \bn \label{e72}&&\hskip-1cm M''M'\int\exp\{i\tint[i\<p,q|\E\,(d/dt)
\E\,|p,q\>/\<p,q|\E\,|p,q\>\nonumber\\
    &&\hskip2.3cm-\<p,q|\E\,{\cal H}\E\,|p,q\>/\<p,q|\E\,|p,q\>]\,dt\}
\,\D_E\mu(p,q)\;.\label{e70}\en
This last relation concludes our second route of calculation beginning 
with (\ref{e66}).

The third relation we wish to derive uses an integral representation for 
the projection operator $\E\,$ generally given by
  \bn \E\,=\tint e^{-i\xi^a\Phi_a(P,Q)}\,f(\xi)\,\delta\xi  \label{e71}\en
for a suitable function $f$. Thus we rewrite (\ref{e66}) in the form
\bn \label{e74}&&\hskip-1.5cm\lim\int\<p'',q''|e^{-i\epsilon\l^a_{N}
\Phi_a}e^{-i\epsilon{\cal H}}e^{-i\epsilon\l^a_{N-1}\Phi_a}
e^{-i\epsilon{\cal H}}\cdots e^{-i\epsilon\l^a_1\Phi_a}
e^{-i\epsilon{\cal H}}e^{-i\epsilon\l^a_0\Phi_a}|p',q'\>\nonumber\\
&&\hskip2cm \times\,f(\epsilon\l_{N})\cdots f(\epsilon\l_0)\,
\delta\epsilon\l_{N}\cdots\delta\epsilon\l_0\;.  \en
Next we insert the coherent-state resolution of unity at appropriate 
places to find that (\ref{e74}) may also be given by
 \bn &&\hskip-1.5cm\lim\int\<p_{N+1},q_{N+1}|e^{-i\epsilon\l^a_{N}
\Phi_a}|p_N,q_N\>\prod_{l=0}^{N-1}\<p_{l+1},q_{l+1}|
e^{-i\epsilon{\cal H}}e^{-i\epsilon\l^a_{l}\Phi_a}|p_l,q_l\>\nonumber\\   
&&\times [\prod_{l=1}^{N}d\mu(p_l,q_l)\,f(\epsilon\l_l)\,\delta\epsilon\l_l]
\,f(\epsilon\l_0)\,\delta\epsilon\l_0\;.  \en
Following the normal pattern, this last expression may readily be turned 
into a formal coherent-state path integral given by
\bn \int\exp\{i\tint[p_j{\dot q}^j-H(p,q)-\l^a(t)\phi_a(p,q)]\,dt\}\,
\D\mu(p,q)\D E(\l)\;,  \en
where $E(\l)$ is a measure designed so as to insert the projection 
operator $\E\,$ at every time slice.  This usage of the Lagrange 
multipliers to ensure that the quantum system remains within the 
quantum constraint subspace is similar to their usage in the classical 
theory to ensure that the system remains on the classical constraint 
hypersurface.  On the other hand, it is also possible to use the measure 
$E(\l)$ in the case of closed {\it first}-class constraints as well; this 
would be just one of the acceptable choices for the measure $C(\l)$ 
designed to put at least one projection operator $\E\,$ into the propagator.

In summary, we have established the equality of the three expressions
\bn &&\hskip-1cm\<p'',q''|\E\;e^{-iT(\E\;{\cal H}\E\;)}\E\,|p',q'\>
\nonumber\\
&&=M''M'\int\exp\{i\tint[i\<p,q|\E\,(d/dt)\E\,|p,q\>/\<p,q|\E\,|p,q\>
\nonumber\\
    &&\hskip2.3cm-\<p,q|\E\,{\cal H}\E\,|p,q\>/\<p,q|\E\,|p,q\>]\,dt\}\,
\D_E\mu(p,q)\nonumber\\
&&= \int\exp\{i\tint[p_j{\dot q}^j-H(p,q)-\l^a(t)\phi_a(p,q)]\,dt\}\,
\D\mu(p,q)\D E(\l)\;.  \en
This concludes our initial derivation of path integral formulas for general 
constraints. Observe that we have not introduced any $\delta$-functionals, 
nor, in the middle expression, reduced the number of integration variables 
or the limits of integration in any way even though in that expression the 
integral over the Lagrange multipliers has been carried out. 

\subsection{Universal procedure to generate single regularized constraints}
The preceding section developed a functional integral approach suitable for
a general set of constraints, but it had one weak point, namely, it required 
prior
knowledge of the constraints themselves in order to choose $f(\xi)$ in 
(\ref{e71}) so as to
construct the appropriate projection operator. Is there any way to construct 
$\E$
{\it without} prior knowledge of the form the constraints will take? The 
answer is {\it yes!}

We first observe that the evolution operator appearing in (\ref{e40}) may 
be written in the 
form of a lattice limit given by
\bn \lim_{\e\ra0}\prod_{1\le n\le N}^{\longleftarrow} \bigg[\T 
e^{-i\tint_{(n-1)\e}^{n\e}\,\H(t)\,dt}
\bigg]\,\bigg[\T e^{-i\tint_{(n-1)\e}^{n\e}\,\l^a(t)\Phi_a\,dt}\bigg]\;, \en
where $\e\equiv T/N$ and the directed product (symbol $\longleftarrow$) 
also respects the time ordering. Thus, this expression is simply an 
alternating sequence of short-time evolutions, first by $\l^a(t)\Phi_a$, 
second by $\H(t)$, a pattern which is then repeated $N-1$ more times. The 
validity of this Trotter-product form follows whenever 
$\H(t)^2+\Phi_a\delta^{ab}\Phi_b$ is essentially self adjoint 
for all $t$, $0\le t\le T$. As a slight generalization, we shall assume 
that $\H(t)^2+\Phi_a M^{ab}\Phi_b$ is essentially self adjoint 
for all $t$, $0\le t\le T$. Here the real, symmetric coefficients 
$M^{ab}\,(=M^{ba})$ are the elements of a positive-definite 
matrix, i.e., $\{M^{ab}\}>0$. For a finite number of constraints, 
$A<\infty$, it is sufficient to assume that $M^{ab}=\delta^{ab}$. 
Other choices for $M^{ab}$ may be relevant when $A=\infty$. 
(We do not explicitly consider the case $A=\infty$ in this article; for some
examples see \cite{klagra}.) 

With all this in mind, we shall explain the construction of a formal 
integration procedure \cite{uni} whereby
\bn  \int \T e^{-i\tint_{(n-1)\e}^{n\e}\,\l^a(t)\Phi_a\,dt}\,\D R(\l)=
\E(\Phi_a M^{ab}\Phi_b\le \delta(\hbar)^2)\;, \en
and for which the integral represented by $\tint\cdots\D R(\l)$ is 
independent of the set of operators $\{\Phi_a\}$ and the Hamiltonian 
operator $\H(t)$ for all $t$. 
First, introduce a formal Gaussian measure $\D S_{\gamma_n}(\l)$ such that
\bn && \hskip-1cm\int\,\T e^{-i\tint_{(n-1)\e}^{n\e}\,\l^a(t)\Phi_a\,dt}\,
\D S_{\gamma_n}(\l)  \nonumber\\
  &&\hskip.35cm={\cal N}\int\,\T e^{-i\tint_{(n-1)\e}^{n\e}\,\l^a(t)\Phi_a
\,dt}\,
e^{(i/4\gamma_n)\tint_{(n-1)\e}^{n\e}\,\l^a(t)(M^{-1})_{ab}\l^b(t)\,dt}\,
\Pi_a\D \l^a\nonumber\\
    &&\hskip.35cm= e^{-i\e\gamma_n(\Phi_aM^{ab}\Phi_b)}\;. \en
The second and last step in the construction involves an 
integration over $\gamma_n$ given by
  \bn && \hskip-1cm\int e^{-i\e\gamma_n(\Phi_a M^{ab}\Phi_b)}
\,d\Gamma(\gamma_n)\nonumber\\ 
&& \hskip.5cm\equiv\lim_{\zeta\ra0^+}\lim_{L\ra\infty}\,
\int_{-L}^L\,e^{-i\e\gamma_n(\Phi_a M^{ab}\Phi_b)}\,
\frac{\sin[\e(\delta^2+\zeta)\gamma_n]}{\pi\gamma_n}\,d\gamma_n\nonumber\\
&&\hskip.5cm=\E(\e\,\Phi_a M^{ab}\Phi_b\le\e\,\delta^2)\nonumber\\
 &&\hskip.5cm=\E(\Phi_a M^{ab}\Phi_b\le\delta^2)\;, \label{e79}\en
which achieves our goal. We note that if 
the final limit is replaced by $\lim_{\zeta\ra0^-}$, the result 
becomes $\E(\Phi_\a M^{\a\beta}\Phi_\beta<\delta^2)$. We normally symbolize 
the pair 
of operations by $\tint\cdots\D R(\l)$, leaving the integral 
over $\gamma_n$ implicit. 

{\bf Remark:} For notational simplicity throughout this article, we generally 
let 
 \bn  &&\int e^{-i\gamma X^2}\frac{\sin(\delta^2\gamma)}{\pi\gamma}
\,d\gamma\nonumber\\ 
&&\hskip1cm \equiv\lim_{\zeta\ra0^+}\lim_{L\ra\infty}\,
\int_{-L}^L\,e^{-i\gamma X^2}\,
\frac{\sin[(\delta^2+\zeta)\gamma]}{\pi\gamma}\,d\gamma\nonumber\\
&&\hskip1cm=\E(X^2\le\delta^2)\;.  \en

With (\ref{e79}) we have found a single, universal procedure to
create the regularized projection operator $\E$ from the set of constraint 
operators in a manner that is
{\it completely independent of the nature of the constraints themselves}.

\subsection{Basic second-class constraint example}
Consider the two degree of freedom system determined by
\bn  I=\tint[p{\dot q}+r{\dot s}-H(p,q,r,s)-\l_1r-\l_2s]\,dt\;, \en
where we have called the variables of the second degree of freedom $r,s$, 
and $H$ is not specified further. The coherent states satisfy $|p,q,r,s\>=
|p,q\>\otimes|r,s\>$, which will be useful. We adopt (\ref{e72}) as our 
formal path integral in the present case, and choose \cite{klasud}
  \bn   &&\E\,=\tint e^{-i(\xi_1R+\xi_2S)}\,e^{-(\xi_1^2+\xi_2^2)/4}\,
d\xi_1d\xi_2/(2\pi)\nonumber\\  
&&\hskip.5cm=\E(R^2+S^2\le\hbar)\equiv|0_2\>\<0_2|  \en
which is a projection operator onto the fiducial vector for the second 
(constrained) degree of freedom only. With this choice it follows that
 \bn &&i\<p,q,r,s|\E\,(d/dt)\E\,|p,q,r,s\>/\<p,q,r,s|\E\,|p,q,r,s\>\nonumber\\
&&\hskip2cm=i\<p,q|(d/dt)|p,q\>-\Im(d/dt)\ln[\<0_2|r,s\>]\nonumber\\
&&\hskip2cm=p{\dot q}-\Im(d/dt)\ln[\<0_2|r,s\>]\;,  \en
and 
\bn &&\hskip-1cm\<p,q,r,s|\E\,\H(P,Q,R,S)\E\,|p,q,r,s\>/\<p,q,r,s|\E\,
|p,q,r,s\>\nonumber\\  &&\hskip1cm=\<p,q,0,0|\H(P,Q,R,S)|p,q,0,0\>\nonumber\\
&&\hskip1cm=H(p,q,0,0)\;. \en
Consequently, for this example, (\ref{e70}) becomes
 \bn {\cal M}\int\exp\{i\tint[p{\dot q}-H(p,q,0,0)]\,dt\}\,\D p\,\D q
\;\times\;\<r'',s''|0_2\>\<0_2|r',s'\>\;, \label{e85}\en
where we have used the fact that at every time slice
 \bn \tint\<r,s|\E\,|r,s\>\,dr\,ds/(2\pi)=\tint|\<0_2|r,s\>|^2\,dr\,ds/(2\pi)
=1\;. \en

Observe, in this path integral quantization, that no variables have been 
eliminated nor has any domain of integration been reduced; moreover, the 
operators $R$ and $S$ have remained unchanged. Also observe that the 
result in (\ref{e85}) is clearly a product of two distinct factors. The 
first factor describes the true dynamics as if we had solved for the 
classical constraints and substituted $r=0$ and $s=0$ in the classical 
action from the very beginning, while the second factor characterizes a 
one-dimensional Hilbert space for the second degree of freedom. Thus we 
can also drop the second factor completely as well as all the integrations 
over $r$ and $s$ and still retain the same physics. In this manner we 
recover the standard result without the use of Dirac brackets or having 
to initially eliminate the second-class 
constraints from the theory.

\subsection{Conversion method}
One common method to treat second-class constraints is to convert them
to first-class constraints and to follow the available procedures for such 
systems; see, e.g., \cite{z2}. Let us first argue classically, and take as 
an example a single degree of freedom with canonical variables $p$ and $q$, 
a vanishing Hamiltonian, and the second-class constraints $p=0$ and $q=0$. 
This situation may be described by the classical action
  \bn  I=\tint[p{\dot q}-\l p-\xi q]\,dt  \;, \en
where $\l$ and $\xi$ denote Lagrange multipliers. Next, let us introduce a 
second canonical pair, say $r$ and $s$, and adopt the classical action
  \bn I'=\tint[p{\dot q}+r{\dot s}-\l(p+r)-\xi(q-s)]\,dt\;. \en
Now the two constraints read $p+r=0$ and $q-s=0$ with a Poisson bracket 
$\{p+r,q-s\}=0$, characteristic of first-class constraints. We obtain the 
original problem by imposing the (consistent) {\it gauge conditions} that 
$r=0$ and $s=0$.
Let us look at this example from the projection operator, coherent state 
approach. 

In the first version with one pair of variables, we are led to the 
reproducing kernel
\bn &&\<p'',q''|\E(P^2+Q^2\le \hbar)|p',q'\>\nonumber\\
&&\hskip1.5cm= \<p'',q''|0\>\<0|p',q'\> \nonumber\\
  &&\hskip1.5cm =e^{-\quarter(p''^2+q''^2-2ip''q'')}\,e^{-\quarter(p'^2
+q'^2+2ip'q')} \;,\label{e90} \en
which provides a ``bench mark'' for this example. As expected the result 
is a one-dimensional Hilbert space.

In the second version of this problem, we start with the expression
\bn  \<p'',q'',r'',s''|\E((P+R)^2+(Q-S)^2\le\delta^2)|p',q',r',s'\>  \en
which involves a constraint with zero in the continuous spectrum. Therefore, 
following previous examples, we multiply this expression with a
suitable factor $c_\delta$ and take the limit as $\delta\ra0$. This factor 
can be chosen so that 
 \bn&& \lim_{\delta\ra0}\,c_\delta\,\<p'',q'',r'',s''|\E((P+R)^2+(Q-S)^2\le
\delta^2)|p',q',r',s'\> \nonumber\\
  &&\hskip1cm =e^{-\quarter[(p''+r'')^2+(q''-s'')^2]+\half i(p''-r'')
(q''-s'')}\nonumber\\
  && \hskip1.5cm\times e^{-\half i(p'-r')(q'-s')-\quarter[(p'+r')^2+
(q'-s')^2]} \;,  \en
an expression which also describes a one-dimensional Hilbert space. This is 
a different (but equivalent) 
representation for the one-dimensional Hilbert space than the one found 
above. Since it is only one-dimensional we
can reduce this reproducing kernel even further, in the fashion illustrated 
earlier, by choosing a ``gauge''
where $r''=s''=r'=s'=0$. When this is done the result becomes 
  \bn e^{-\quarter(p''^2+q''^2-2ip''q'')}\,e^{-\quarter(p'^2+q'^2+2ip'q')} 
\;, \en
which is {\it identical} to the expression (\ref{e90}) found by quantization 
of the second-class constraints directly. In this manner we see how the 
conversion method, in which second-class constraints are turned into 
first-class constraints by the introduction of auxiliary degrees of 
freedom, appears within the projection operator, coherent state approach 
as well. Applications of the conversion method made within the projection 
operator approach may be found in \cite{sha8}.

\subsection{Equivalent representations}
In dealing with quantum mechanics, one may employ many different---yet 
equivalent---representations of the vectors and operators involved.
While, in certain circumstances,  some representations may be more 
convenient than others, the notion that some representations are 
``better'' than others should be resisted.

In the context of coherent-state representations, for example, a change 
of the fiducial vector leads to an equivalent representation. If, for
a rather general (normalized) fiducial vector $|\eta\>$,  we set
  \bn  |p,q;\eta\>\equiv e^{-iqP}\,e^{ipQ}\,|\eta\>\;,  \en
then
  \bn  \psi(p,q;\eta)\equiv\<p,q;\eta|\psi\>  \en
defines $\eta$-dependent representatives of the abstract vector $|\psi\>$. 
However, all 
representation-dependent aspects disappear when
physical questions are asked such as
  \bn  \tint |\psi(p,q;\eta)|^2\,(dp\,dq/2\pi)=\<\psi|\psi\>\;.  \en

More general representation issues may be addressed by using arbitrary 
unitary operators, say $V$. Thus if $|p,q\>$ denotes elements of one
(say) coherent state basis, then $|p,q;V\>\equiv V^\dagger|p,q\>$ denotes 
the elements of another basis. Vector and operator
representatives, $\psi(p,q;V)\equiv \<p,q;V|\psi\>$ and $A(p',q';V:p,q;V)
\equiv\<p',q';V|{\cal A}|p,q;V\>$, respectively, provide equivalent sets 
of functional representatives for different $V$. Evidently the {\it physics} 
is unchanged in this transformation; only the intermediate {\it 
mathematical representatives} are affected. This formulation is similar 
to passive coordinate transformations in other disciplines. 
Another version similar to active coordinate transformations is also 
possible. In this version the basis vectors, say $|p,q\>$, for all relevant 
$(p,q)$, remain unchanged; instead, the abstract vectors $|\psi\>$ and 
operators ${\cal A}$, etc., are transformed: $|\psi\>\ra V|\psi\>$, 
${\cal A}\ra V{\cal A}V^\dagger$, etc. It is this form of equivalence that 
we turn to next.

\subsection{Equivalence of criteria for \\second-class constraints}
Let us return to the simple example of second-class constraints discussed 
above where, classically, $p=q=0$. In the associated quantum theory, we 
chose to express these constraints with the help of the projection operator 
$\E=\E(P^2+Q^2\le\hbar)=|0\>\<0|$, namely, the projection operator onto 
the ground state of the ``Hamiltonian'' $P^2+Q^2$. In turn, this expression 
led directly to the coherent-state representation of $\E$ given by 
$\<p',q'|\E|p,q\>=
\<p',q'|0\>\<0|p,q\>$. However, the question arises, what is special 
about the combination $P^2+Q^2$? As we shall now argue, any other possible 
choice leads to an equivalent representation.

As a first example, consider
  \bn  \E(P^2+\omega^2 Q^2\le \omega\hbar)=|0;\omega\>\<0;\omega|=
V^\dagger_\omega\,|0\>\<0|\,V_\omega \;, \en
where $V_\omega$ denotes a suitable unitary operator, 
which establishes the equivalence for any $\omega$, $0<\omega<\infty$. We 
emphasize that we do {\it not} assert the unitary equivalence of
$P^2+Q^2$ and $P^2+\omega^2 Q^2$ for any value of $\omega\ne1$, only that 
$|0;\omega\>$ and $|0\>$ are unitarily related---as are {\it any} two unit 
vectors in Hilbert space.

Furthermore, there is nothing sacred about the quadratic combination. For 
example, for any $0<\l<\infty$, consider $\E(P^2+\l 
Q^4\le\delta(\hbar)^2)\equiv|0,\lambda\>\<0,\lambda|$, where we have 
adjusted $\delta(\hbar)$ to the lowest eigenvalue so as to include only 
a single eigenvector, $|0,\lambda\>$. Since there exists a unitary 
operator $V_\l$ such that $\<0,\lambda|=\<0|V_\l$, this choice of 
projection operator leads to an equivalent coherent-state
representation as well.

More generally, we are led to reconsider the projection operator
  \bn   \E(\Sigma_a\Phi^2_a\le\delta(\hbar)^2)=\sum_{j=1}^J\,|j\>\<j|\;, 
\label{e98} \en
where $\<j|k\>=\delta_{jk}$ and $1\le J\le\infty$, as determined by the 
choice of $\delta(\hbar)$. Since all $J$-dimensional subspaces are
unitarily equivalent to each other (with suitable care taken when 
$J=\infty$), the given prescription is entirely equivalent to any other 
version, such as 
  \bn  \E({\cal F}(\Phi_a)\le {\tilde\delta}(\hbar)^2)=
\sum_{{\sf j}=1}^{\sf J}\,|{\sf j}\>\<{\sf j}|\;, \label{e99} \en
where $\<{\sf j}|{\sf k}\>=\delta_{{\sf j}{\sf k}}$, provided that 
${\tilde\delta}(\hbar)$ may be---and is---chosen so that ${\sf J}=J$. 
Here ${\cal F}(\Phi_a)$ denotes a nonnegative self-adjoint operator that 
includes all the constraint operators, and for very small 
${\tilde\delta}(\hbar)^2$ forces the spectral contribution of each 
constraint operator to be correspondingly small, just as is the case in 
(\ref{e98}).

In summary, the general, quadratic criterion we have adopted in (\ref{e98}) 
has been chosen for simplicity and convenience; any other restriction on the
constraint operators leads to an equivalent theory, as in (\ref{e99}), 
provided that the dimensionality of $\E$ remains the same.

\section{SELECTED EXAMPLES OF \\FIRST-CLASS CONSTRAINTS}
\subsection{General configuration space geometry}
Although we shall discuss constraints that lead to a general configuration
 space geometry in this section, we shall for the most part use rather
 simple illustrative examples. To begin with let us consider the constraint
 \bn   \sum_{j=1}^J(q^j)^2=1\;,  \en
a condition which puts the classical problem on a (hyper)sphere of unit
 radius. For convenience in what follows we shall focus as well on the case
 of a vanishing Hamiltonian so as to isolate clearly the consequences of the
 constraint independently of any dynamical effects. Adopting a standard 
vector inner product 
notation and a different kinematic term, 
 consider the formal path integral
  \bn {\cal M}\int\exp\{i\tint[-q\cdot{\dot p}-\l(q^2-1)]\,dt\}\,\D p\,\D
 q\,\D C(\l)\;,  \label{f100}\en
the result of which is given by
  \bn  \<p'',q''|\E\,|p',q'\> \en
where
  \bn  \E=\int_{-\infty}^\infty
 e^{-i\l(Q^2-1)}\,\frac{\sin(\delta\l)}{\pi\l}\,d\l 
  =\E\,(-\delta\le Q^2-1\le\delta)\;.  \en

In order, ultimately, to obtain a suitable reduction of the reproducing 
kernel in the
 present case, we  allow for fiducial vectors other than harmonic oscillator
 ground states. Thus we let $|\eta\>$ denote a general unit vector for the
 moment; its required properties will emerge from our analysis. In 
accordance with (\ref{f100}), we choose a
 phase convention for the coherent states---in particular, in (\ref{e17}) 
we multiply
 by $e^{ip{\cdot}q}$---so that now the Schr\"odinger
 representation of the coherent states reads
 \bn \<x|p,q\>=e^{ip\cdot x}\,\eta(x-q)\;,  \en
which leads immediately to the expression
  \bn \<p'',q''|p',q'\>=\int\eta^*(x-q'')\,e^{-i(p''-p')\cdot
 x}\,\eta(x-q')\,d^J\!x\;. \en
Consequently, the reproducing kernel that incorporates the projection
 operator is given, for $0<\delta<1$, by
  \bn
 \<p'',q''|\E\,|p',q'\>=\int_{1-\delta\le x^2\le 1+\delta}\eta^*(x-q'')
\,e^{-i(p
''-p')\cdot x}\,\eta(x-q')\,d^J\!x\;. \label{e106}\en
Since $\E\,$ represents a projection operator, it is evident that this
 expression defines a reproducing kernel which admits a local integral for
 its inner product (for any normalized $\eta$) with a measure
 $d^J\!p\,d^J\!q/(2\pi)^J$ and an integration domain $\ir^{2J}$.

However, if we are willing to restrict our choice of fiducial vector, we can
 reduce the number of integration variables and change the domain of
 integration in a meaningful way. Recall that the group ${\rm E}(J)$, the
 Euclidean group in $J$-dimensions, consists of rotations that preserve the
 unit (hyper)sphere in $J$-dimensions, as well as $J$ translations. As
 emphasized by Isham \cite{ish}, this is the natural canonical group for a
 system confined to the surface of a (hyper)sphere in $J$ dimensions. We can
 adapt our present coherent states to be coherent states for the group ${\rm
 E}(J)$ without difficulty.

To that end consider the reduction of the reproducing kernel (\ref{e106}) 
to one for
 which $q''^2=q'^2\equiv1$. To illustrate the process as clearly as
 possible let us choose $J=2$. As a consequence we introduce
  \bn 
 \<a'',b'',c''|a',b',c'\>\equiv\<p'',q''|\E\,|p',q'\>_{q''^2=q'^2=1}\;,
 \en
where $a\equiv p_1$, $b\equiv p_2$, and $c$ arises from the identification
 $q^1\equiv\cos(c)$ and $q^2\equiv\sin(c)$, all relations holding for both
 end points. Expressed in terms of polar coordinates, $r,\phi$, the reduced
 reproducing kernel becomes
  \bn &&\hskip-1.3cm\<a'',b'',c''|a',b',c'\> \nonumber \\  
 &&\hskip-1.1cm=\int_{|r^2-1|\le\delta}\!\eta^*(r,\phi-c'')\,
e^{-i(a''-a')r\cos
\phi-i(b''-b')r\sin\phi}\,\eta(r,\phi-c')\,r\,dr\,d\phi\,.  \en

We next seek to choose $\eta$, if at all possible, in such a way that the
 inner product of this new (reduced) reproducing kernel admits a local
 integral for its inner product. As a starting point we choose the
 left-invariant group measure for ${\rm E}(2)$ which is given by
 $M\,da\,db\,dc$, $M$ a constant, with an integration domain $\ir^2\times
 S^1$. Therefore, we are led to propose that
 \bn&&\hskip-1cm\int\int_{|r^2-1|<\delta}\eta^*(r,\phi-c'')\,e^{-i(a''-a)r
\cos\phi-i(b''-b)r\sin\phi}\,\eta(r,\phi-c)\,r\,dr\,d\phi\nonumber\\
&&\hskip-.1cm\times\int_{|\rho^2-1|<\delta}\eta^*(\rho,\theta-c)\,e^{-i(a-a'
)\rho\cos\theta-i(b-b')\rho\sin\theta}\,\eta(\rho,\theta-c')\,\rho\,d\rho\,d
\theta\nonumber\\
&&\hskip-.1cm\times M\,da\,db\,dc  \nonumber\\
&&\hskip-.5cm=(2\pi)^2M\int\eta^*(r,\phi-c'')e^{-i(a''-a')r\cos\phi-i(b''-
b')r\sin\phi}\,\eta(r,\phi-c')\,r\,dr\,d\phi\nonumber\\
&&\hskip-.1cm\times\int|\eta(r,c)|^2\,dc\;,  \en
which leads to the desired result provided (i)
  \bn  \int_0^{2\pi}|\eta(r,c)|^2\,dc=P\;,\hskip1cm 0<P<\infty\;,  
\label{e110}\en
is {\it independent} of $r$, $|r^2-1|<\delta$, and (ii)
 $M=[(2\pi)^2\,P]^{-1}$. Given a general nonvanishing vector $\xi(r,\phi)$,
 a vector satisfying (\ref{e110}) may always be given by
  \bn 
 \eta(r,\phi)=\xi(r,\phi)/\sqrt{\tint_0^{2\pi}|\xi(r,\theta)|^2\,d\theta}
 \en   provided the denominator is positive, and
which specifically leads to $P=1$. In this way we have reproduced the ${\rm
 E}(2)$-coherent states of Ref.~\cite{ikl}, even including the necessity for
 a small interval of integration in $r$, and where fiducial vectors
 satisfying (\ref{e110}) were called ``surface constant''.

Dynamics consistent with the constraint $q^2=1$ is obtained in the ${\rm
 E}(2)$ case by choosing a Hamiltonian that is a function of the coordinates
 on the circle, namely $\cos(\theta)$ and $\sin(\theta)$, as well as the
 rotation generator of ${\rm E}(2)$, i.e., $-i\d/\d\theta$. We refer the
 reader to \cite{ikl} for a further discussion of ${\rm E}(2)$-coherent
 states as well as a discussion of the introduction of compatible dynamics.
 An analogous discussion can be given for the classical constraint $q^2=1$
 for any value of $J>2$.

Not only can compact (hyper)spherical configuration spaces be treated in
 this way, but one may also treat noncompact (hyper)pseudospherical spaces
 defined by the constraint
  \bn \Sigma_{i=1}^Iq^{i\,2}-\Sigma_{j=I+1}^Jq^{j\,2}=1\;,\hskip1cm1
\leq I\leq J-1\;,  \en
appropriate to the Euclidean group ${\rm E}(I,J-I)$. Such an analysis would 
lead
 to ${\rm E}(I,J-I)$-coherent states.

Finally, we comment on the constraint of a general curved configuration
 space which can be defined by a set of compatible constraints
 $\phi_a(q)=0$.
Clearly these constraints satisfy $\{\phi_a(q),\phi_b(q)\}=0$, and define
 a $(J-A)$-dimensional configuration space in the original Euclidean
 configuration space $\ir^J$. The relevant projection operator
 $\E\,=\E\,(\Sigma\Phi_a^2(Q)\le\delta^2)$ is defined in an evident fashion,
 and the reproducing kernel incorporating the projection operator is defined
 in analogy with the prior discussion. This reproducing kernel enjoys a
 local integral representation for its inner product, in fact, this integral 
is with
 the same measure and integration domain as without the projection operator.
 What differs in the present case is that when the reproducing kernel is put
 on the constraint manifold, the resultant coherent states are generally
 {\it not} defined by the action of a group on a fixed fiducial vector. In
 short, the relevant coherent states are not group generated, which, in fact, 
is consistent with their most basic definition; see, e.g., \cite{klska,kl12}.

\subsection{Finite-dimensional Hilbert space examples}
Let us consider the case of two degrees of freedom with a ``classical''
 action function given by
\bn I =\tint[\half(p_1{\dot q}_1-q_1{\dot p}_1+p_2{\dot q}_2-q_2{\dot
 p}_2)-\l(p_1^2+p_2^2+q_1^2+q_2^2-4s\hbar)]\,dt  \en
For clarity of presentation, we explicitly include $\hbar$
 in our classical action, and we continue to make it explicit it throughout 
this 
section. With the present phase convention 
for the coherent states, the
 unconstrained reproducing kernel is given by
 \bn &&\<p'',q''|p',q'\>\equiv\<z''|z'\> \nonumber\\
&&\hskip2.3cm=\exp[\Sigma_{j=1}^2(-\half|z''_j|^2+{z''^*}_jz'_j-\half|z'
_j|^2)]  \en
where $z_j\equiv (q_j+ip_j)/\sqrt{2\hbar}$ for each of the end points.

We next observe that the constraint operator 
 \bn \Phi=:P^2_1+P^2_2+Q^2_1+Q^2_2:-4s\hbar\one  \en
has discrete eigenvalues, i.e., $2(n_1+n_2-2s)\hbar$, where $n_1$ and $n_2$
 are nonnegative integers, based on the choice of $|\eta\>$ as the ground
 state for each oscillator. To satisfy $\Phi=0$ it is necessary that $2s$
 be an integer in which case the quantum constraint subspace is
 $(2s+1)$-dimensional. The projection operator in the present case is
 defined by
 \bn  \E\,=\pi^{-1}\int_0^{\pi}
 \exp[\,-i\l(:P^2_1+P^2_2+Q^2_1+Q^2_2:-4s\hbar\one)/\hbar]\,d\l  \en
which projects onto the appropriate $(2s+1)$-dimensional subspace. 
It is straightforward to
 demonstrate that
  \bn
 &&\hskip-1.3cm\<z''|\E\,|z'\>=\exp[-\half\Sigma_{j=1}^2(|z''_j|^2+|z'_j
|^2)] {[(2s)!]}^{-1}({z''^*}_1z'_1+{z''^*}_2z'_2)^{2s}\nonumber\\
&&\hskip-1.2
cm=\exp[-\half\Sigma_{j=1}^2(|z''_j|^2+|z'_j|^2)]
 \sum_{k=0}^{2s}[{k!(2s-k)!}]^{-1}({z''^*}_1z'_1)^k({z''^*}_2z'_2)^{2s-k}
  \en
The projected reproducing kernel in this case corresponds to a {\it finite}
 dimensional Hilbert space; nevertheless, the inner product is given by the
 same measure and integration domain as in the original, unprojected,
 infinite dimensional Hilbert space!

Of course, there are other, simpler and more familiar ways to represent a
 finite-dimensional Hilbert space; but any other representation is evidently
 equivalent to the one described here.

As the notation suggests the present quantum constraint subspace provides a
 natural carrier space for an irreducible representation of $\rm SU(2)$ with
 spin $s$.  We observe that the following three expressions represent
 generators of the classical rotation group in their action on the constraint
 hypersurface:
 \bn &&s_x=\half(p_1p_2+q_1q_2)\;,  \nonumber\\
     &&s_y=\half(q_1p_2-p_1q_2)\;,  \nonumber\\
     &&s_z=\quarter(p_1^2+q_1^2-p^2_2-q_2^2)\;.   \en
Thus these quantities serve as potential ingredients for a Hamiltonian which
 is compatible with the constraint. 

Although not the subject of this section, we may also observe that an
 analogous discussion holds in case of the constraint
\bn \phi(p,q)=p_1^2+q_1^2-p_2^2-q_2^2-2k\hbar=0\;,   \en
where $k$ is an integer, and the resultant reduced Hilbert space is infinite
 dimensional for any integral $k$ value. In this case the relevant group is 
SU(1,1).
\subsection{Helix model}
In \cite{lee}, Friedberg, Lee, Pang, and Ren analyzed the so-called helix
 model. For details of
 this model (see also \cite{z3}) and its possible role as a simple analogue 
of the Gribov problem 
 in non-Abelian gauge
 models, we refer the reader to their paper. We begin with the classical
 Hamiltonian for a three-degree of freedom system given by
\bn H=\half(p_1^2+p_2^2+p_3^2)+U(q_1^2+q_2^2)+\l[g(p_2q_1-q_2p_1)+p_3]\
,\en
where $U$ denotes the potential, which hereafter, following \cite{lee}, we
 shall choose as harmonic, namely $U(q_1^2+q_2^2)=\omega^2(q_1^2+q_2^2)/2$
, because then this special model is fully soluble. Here, $g>0$ is a
 coupling constant, and $\l=\l(t)$ is the Lagrange multiplier which
 enforces the single first-class constraint
 \bn \phi(p,q)=g(p_2q_1-q_2p_1)+p_3=0\;.  \en

For the first two degrees of freedom we choose coherent states with the
 phase convention adopted for the previous example, while for the third
 degree of freedom we return to the original phase convention. This choice 
means
 that we consider the formal coherent state path integral given by
  \bn &&\hskip-1cm\int\exp(\!\!(i\tint\{\half(p_1{\dot q}_1-q_1{\dot
 p}_1)+\half(p_2{\dot q}_2-q_2{\dot p}_2)+p_3{\dot
 q}_3\nonumber\\&&\hskip.8cm-\half(p_1^2+p_2^2+p_3^2)-\half\omega^2(q_1^2+q_
2^2)\nonumber\\&&\hskip.8cm-\l[g(p_2q_1-q_2p_1)+p_3]\}\,dt)\!\!)\,
\D\mu(p,q)\,\D
 C(\l)\nonumber\\
   &&=\<z''_1,z''_2,p''_3,q''_3|\,e^{-i{\cal
 H}T}\,\E\,|z'_1,z'_2,p'_3,q'_3\>\;.  \en
In the present case the relevant projection operator $\E\,$ is given (for
 $\hbar=1$, and $0<\delta\ll g$) by
 \bn
 \E\,=\E\,((gL_3+P_3)^2\le\delta^2)=\sum_{m=-\infty}^\infty\E\,((gm+P_3)^2\le
\delta^2)\,\E\,(L_3=m)\;,  \en
where we have used the familiar spectrum for the rotation generator $L_3$.
If ${\cal H}_0$ denotes the harmonic oscillator Hamiltonian for the first
 two degrees of freedom, then it follows that
\bn &&\hskip-.3cm\<z''_1,z''_2,p''_3,q''_3|\,e^{-i{\cal
 H}T}\,\E\,|z'_1,z'_2,p'_3,q'_3\>\nonumber\\
 &&=\sum_{m=-\infty}^\infty\<z''_1,z''_2|e^{-i{\cal
 H}_0T}\E\,(L_3=m)\,|z'_1,z'_2\>\nonumber\\
 &&\hskip.3cm\times\<p''_3,q''_3|e^{-iP_3^2T/2}\E\,(-\delta\le gm+P_3
\le\delta)|p
'_3,q'_3\>\nonumber\\
&&=\exp[-\half(|z''_1|^2+|z''_2|^2+|z'_1|^2+|z'_2|^2)]\nonumber\\
&&\hskip.3cm\times\sum_{m=-\infty}^\infty\Big\{\frac{(z''^*_1+iz''^*_2)(z'
_1-iz'_2)}{(z''^*-iz''^*_2)(z'_1+iz'_2)}\Big\}^{m/2}\!I_m(\sqrt{(z''^{*2}_1+
z''^{*2}_2)(z'^2_1+z'^2_2)}\,e^{-i\omega T})\nonumber\\
&&\hskip.3cm\times\exp[-\half(gm+p''_3)^2-\half(gm+p'_3)^2-i\half
 g^2m^2T-igm(q''_3-q'_3)]\nonumber\\
&&\hskip.3cm\times\frac{2}{\sqrt{\pi}}\frac{\sin[\delta(q''_3-q'_3)]}{(q''_3
- -q'_3)}+O(\delta^2)\;, \label{e124} \en
where $I_m$ denotes the usual Bessel function.

We observe that the spectrum for the Hamiltonian agrees with the results of
 Ref.~\cite{lee}, and moreover, to leading order in $\delta$,  we have
 obtained gauge-invariant results, i.e., insensitivity to any choice of the
 Lagrange multiplier function $\l(t)$, merely by projecting onto the quantum
 constraint subspace at $t=0$. The constrained propagator (\ref{e124}) is 
composed
 with the same measure and integration domain as is the unconstrained
 propagator.  We
 may also divide the constrained propagator by $\delta$ and take the limit
 $\delta\ra0$. The result is a new functional expression for the propagator
 that fully satisfies the constraint condition, but one that no longer
 admits an inner product with the same measure and integration domain as
 before.

\subsection{Reparameterization invariant dynamics}
Let us start with a single degree of freedom $(J=1)$ and the action
 \bn \tint[p{\dot q}-H(p,q)]\,dt\;.  \en
We next promote the independent variable $t$ to a dynamical variable, 
introduce $s$ as its conjugate momentum (often called $p_t$), enforce the 
constraint $s+H(p,q)=0$, and lastly introduce $\tau$ as a new independent 
variable. This modification is realized by means of the classical action
\bn \tint\{pq^*+st^*-\l[s+H(p,q)]\}\,d\tau\;,  \en
where $q^*=dq/d\tau$, $t^*=dt/d\tau$, and $\l=\l(\tau)$ is a Lagrange 
multiplier. The coherent-state path integral is constructed so that
  \bn
&&\hskip-.5cm{\cal M}\int\exp(\!\!(i\tint\{pq^*+st^*-\l[s+H(p,q)]\}\,dt)
\!\!)\,\D p\,\D q\,\D s\,\D t\,\D C(\l)\nonumber\\
&&\hskip1.5cm=\<p'',q'',s'',t''|\E\,|p',q',s',t'\>\;, \label{e127} \en
where
  \bn  &&\E\,= \int_{-\infty}^\infty e^{-i\xi[S+\H(P,Q)]}\;\frac{
\sin(\delta\xi)}{\pi\xi}\,d\xi\nonumber\\
&&\hskip.41cm=\E\,(-\delta\le S+\H(P,Q)\le\delta)\;.  \label{e128}\en
The result in (\ref{e127}) and (\ref{e128}) represents as far as we can go 
without choosing ${\cal H}(P,Q)$.

To gain further insight into such expressions, we specialize to the case of 
the nonrelativistic free particle, $\H=P^2/2$. Then it follows that
\bn  &&\<p'',q'',s'',t''|\E\,|p',q',s',t'\>\nonumber\\    
  &&\hskip.5cm=\pi^{-1}\int_{-\infty}^\infty\exp[-\half(k-p'')^2-\half(
\half k^2+s'')^2\nonumber\\
  &&\hskip2.85cm+ik(q''-q')-i\half k^2(t''-t')\nonumber\\
  &&\hskip2.85cm-\half(k-p')^2-\half(\half k^2+s')^2]\,dk\nonumber\\  
&&\hskip2.3cm\times\frac{2\sin[\delta(t''-t')]}{(t''-t')}+O(\delta^2)\;.\en
For any $\delta$ such that $0<\delta\ll1$, we observe that this expression 
represents a reproducing kernel which in turn defines an associated 
reproducing kernel Hilbert space composed, as usual, of bounded, continuous 
functions given, for arbitrary complex numbers $\{\alpha_k\}$, phase-space 
points $\{p_k,q_k,s_k,t_k\}$, and $K<\infty$, by
\bn  \psi(p,q,s,t)\equiv\sum_{k=0}^K\alpha_k\<p,q,s,t|\E\,|
p_k,q_k,s_k,t_k\>\;,  \en
or as the limit of Cauchy sequences of such functions in the norm defined 
by means of the inner product given by
\bn (\psi,\psi)=\tint|\psi(p,q,s,t)|^2\,dp\,dq\,ds\,dt/(2\pi)^2  \en
integrated over $\ir^4$. 

Let us next consider the reduction of the reproducing kernel given by 
\bn &&\hskip-1cm\<p'',q'',t''|p',q',t'\>\nonumber\\  
&&\equiv\lim_{\delta\ra0}\frac{1}{4\sqrt{\pi}\,\delta}\int\<p'',q'',s'',t''|
\E\,|p',q',s',t'\>\,ds''\,ds'\nonumber\\
&&=\pi^{-1/2}\int\exp[-\half(k-p'')^2-\half(k-p')^2\nonumber \\
  &&\hskip2.68cm+ik(q''-q')-i\half k^2(t''-t')]\,dk\;,  \en
which in turn generates a {\it new} reproducing kernel in the indicated 
variables. For the resultant kernel it is straightforward to demonstrate, 
for any $t$, that 
 \bn \int\<p'',q'',t''|p,q,t\>\<p,q,t|p',q',t'\>\,dp\,dq/(2\pi) 
=\<p'',q'',t''|p',q',t'\> \label{e133}\;. \en
This relation implies that the span of the vectors $\{|p,q\>\equiv|p,q,0\>\}$ 
is identical with the span of the vectors $\{|p,q,t\>\}$, meaning further
that the states $\{|p,q,t\>\}$ form a set of {\it extended coherent states}, 
which are ``extended'' with respect to $t$ in the sense of Ref.~\cite{kwh}. 
Observe how the time variable has become distinguished by the criterion 
(\ref{e133}). Consequently, we may properly interpret
 \bn \<p'',q'',t''|p',q',t'\>\equiv\<p'',q''|e^{-i(P^2/2)(t''-t')}|
p',q'\>\;, \en
namely, as the conventional, single degree of freedom, coherent-state matrix 
element of the evolution operator appropriate to the free particle. 

To further demonstrate this interpretation as the dynamics of the free 
particle, we may pass to sharp $q$ matrix elements with the observation 
that
\bn &&\hskip-1.5cm\<q''|e^{-i(P^2/2)(t''-t')}|q'\>\nonumber\\
&&\equiv\frac{\pi^{1/2}}{(2\pi)^2}\int\<p'',q''|e^{-i(P^2/2)(t''-t')}|
p',q'\>\,dp''\,dp'\nonumber\\
&&=\frac{1}{2\pi}\int\exp[\,ik(q''-q')-i\half k^2(t''-t')]\,dk\nonumber\\
&&=\frac{\;\;e^{i(q''-q')^2/2(t''-t')}}{\sqrt{2\pi i(t''-t')}}\;, \en
which is clearly the usual result.
\subsection{Elevating the Lagrange multiplier to an \\additional dynamical 
variable}
Sometimes it is useful to consider an alternative formulation of a system 
with constraints in which the initial Lagrange multipliers are regarded as 
dynamical variables, complete with their own conjugate variables, and to 
introduce new constraints as needed. For example, let us start with a single 
degree of freedom system with a single first-class constraint specified by 
the action functional
  \bn \tint[p{\dot q}-H(p,q)-\l\phi(p,q)]\,dt\;,  \en
where $\phi(p,q)$ represents the constraint and $\l$ the Lagrange 
multiplier. Instead, let us replace this action functional by
  \bn \tint[p{\dot q}+\pi{\dot\l}-H(p,q)-\sigma\pi-\theta\phi(p,q)]\,dt\;. \en
In this expression we have introduced $\pi$ as the canonical conjugate to 
$\l$, the Lagrange multiplier $\sigma$ to enforce the constraint $\pi=0$, 
and the Lagrange multiplier $\theta$ to enforce the original constraint 
$\phi=0$. Observe that $\{\pi,\phi(p,q)\}=0$, and therefore the constraints 
remain first class in the new form. The path integral expression for the 
extended form reads
  \bn &&\hskip-.7cm{\cal M}\int\exp\{i\tint[p{\dot q}+\pi{\dot\l}-H(p,q)-
\sigma\pi-\theta\phi(p,q)]\,dt\}\,\D p\,\D q\,\D\pi\,\D\l\,\D C(\sigma,
\theta)\nonumber\\
&&\hskip1.3cm=\<p'',q'',\pi'',\l''|e^{-i{\cal H}T}\E\,|p',q',\pi',\l'\>\;. \en
In this expression, we may choose
  \bn  \E\,=\E\,(\Phi(P,Q)^2\le\delta^2)\,\E\,(\Pi^2\le\delta'^2) \en
involving two possibly distinct reglarization parameters.
Consequently, the complete propagator factors into two terms,
 \bn &&\<p'',q'',\pi'',\l''|e^{-i{\cal H}T}\E\,|p',q',\pi',\l'\>\nonumber\\
&&\hskip2cm=\<p'',q''|e^{-i{\cal H}T}\E\,(\Phi(P,Q)^2\le\delta^2)|p',q'\>
\nonumber\\
&&\hskip2.15cm\times\<\pi'',\l''|\E\,(\Pi^2\le\delta'^2)|\pi',\l'\>\;. \en

The first factor is exactly what would be found by the appropriate path 
integral of the original classical system with only the single constraint 
$\phi(p,q)=0$ and the single Lagrange multiplier $\l$. The second factor 
represents the modification introduced by considering the extended system. 
Note, however, that with a suitable $\delta'$-limit the second factor 
reduces to a product of terms, one depending on the ``$\;''\;$'' arguments, 
the other depending on the ``$\;'\;$'' arguments, just as was the case 
previously. This result for the second factor implies that it has become 
the reproducing kernel for a {\it one-dimensional} Hilbert space, and when 
multiplied by the first factor it may be ignored entirely. In this way it 
is found that the quantization of the original and extended systems leads 
to identical results.


\section{SPECIAL APPLICATIONS}
\subsection{Algebraically inequivalent constraints}
The following example is suggested by Problem 5.1 in Ref.~\cite{hen}. 
Consider the two-degree of freedom system with vanishing Hamiltonian 
described by the classical action
  \bn I=\tint(p_1{\dot q}_1+p_2{\dot q}_2-\l_1p_1-\l_2p_2)\,dt\;.  \en
The equations of motion become 
 \bn {\dot q}_j=\l_j\;,\hskip.8cm{\dot p}_j=0\;,\hskip.8cm p_j=0\;,
\hskip1.2cm j=1,2\;.\en
Evidently the Poisson bracket $\{p_1,p_2\}=0$.

As a second version of the same dynamics, consider the classical action
 \bn  I=\tint(p_1{\dot q}_1+p_2{\dot q}_2-\l_1p_1-\l_2e^{cq_1}p_2)\,dt\;,  \en
which leads to the equations of motion
\bn {\dot q}_1=\l_1\;,\hskip.3cm{\dot q}_2=\l_2e^{cq_1}\;,
\hskip.3cm{\dot p}_1=-c\l_2e^{cq_1}p_2\;,\hskip.3cm{\dot p}_2=0\;,
\hskip.3cm p_1=e^{cq_1}p_2=0\en
Since $e^{cq_1}p_2=0$ implies that $p_2=0$, it follows that the two 
formulations are equivalent despite the fact that in the second case
$\{p_1,e^{cq_1}p_2\}=-c\,e^{cq_1}p_2$, which has a fundamentally different 
algebraic structure when $c\neq0$ as compared to $c=0$.

Let us discuss these two examples from the point of view of a coherent 
state, projection operator quantization. For the first version we consider
\bn {\cal M}\int\exp[i\tint(p_1{\dot q}_1+p_2{\dot q}_2-\l_1p_1-\l_2p_2)
\,dt]\,\D p\,\D q\,\D C(\l)\;,  \en
defined in a fashion to yield
 \bn \<p'',q''|\E|p',q'\>\en
where, for ease of evaluation, we may choose
 \bn  \E=\E(P_1^2\le\delta^2)\E(P_2^2\le\delta^2)\;.  \en
In particular this choice leads to the fact that
 \bn &&\<p'',q''|\E|p',q'\>\nonumber\\
   &&\hskip-1cm=\pi^{-1}\prod_{l=1}^2\int_{-\delta}^\delta 
\exp[-\half(k_l-p''_l)^2+ik_l(q''_l-q'_l)-\half(k_l-p'_l)^2]\,dk_l\;. \en
Let us reduce this reproducing kernel, in particular, by multiplying 
this expression by $\pi/(2\delta)^2$ and passing to the limit $\delta\ra0$. 
The result is the reduced reproducing kernel given by
  \bn \exp[-\half(p''^2_1+p''^2_2)]\,\exp[-\half(p'^2_1+p'^2_2)]\;, \en
which clearly characterizes a particular representation of a 
one-dimensional Hilbert space in which every vector is proportional 
to $\exp[-\half(p_1^2+p_2^2)]$. This example, of course, is related 
to the reduction examples given earlier. Moreover, we can introduce an 
integral representation over the remaining $p$ variables for the inner 
product if we so desire.

Let us now turn attention to the second formulation of the problem by 
focussing [for a different $C(\l)$] on
\bn {\cal M}\int\exp[i\tint(p_1{\dot q}_1+p_2{\dot q}_2-\l_1p_1-\l_2
e^{cq_1}p_2)\,dt]\,\D p\,\D q\,\D C(\l)\;.  \en
This expression again leads (for a different $\E$) to 
 \bn \<p'',q''|\E|p',q'\>\;,\en
where in the present case the fully {\it reduced} form of this expression 
is proportional to
  \bn &&\hskip-1cm\int\exp[-\half(k_2-p''_2)^2+ik_2(q''_2-q'_2)-
\half(k_2-p'_2)^2]\nonumber\\
  &&\times\exp[-\half(k_1-p''_1)^2+ik_1q''_1-\half i\l_1k_1]\nonumber\\
  &&\times\exp[-ixk_1-i\l_2e^{cx}k_2+ix\kappa_1]\nonumber\\
  &&\times\exp[-\half i\l_1\kappa_1-i\kappa_1q'_1-\half(\kappa_1-p'_1)^2]
\nonumber\\
  &&\hskip1cm \times 
dk_2\,dk_1\,dx\,d\kappa_1\,d\l_1\,d\l_2\;.  \en
When normalized appropriately, this expression is evaluated as
  \bn \exp[-\half(p''^2_1+p''^2_2+icp''_1)]\,\exp[-\half(p'^2_1+p'^2_2-
icp'_1)]\;,
\en
which once again represents a one-dimensional Hilbert space although 
it has a different representation than in the case $c=0$. 

Thus we have obtained a $c$-dependent family of distinct but equivalent 
quantum representations for the same Hilbert space, reflecting the 
$c$-dependent family of equivalent classical solutions.
\subsection{Irregular constraints}
In discussing constraints one often pays considerable attention to the 
regularity of the expressions involved. Consider, once again, the simple 
example of a single constraint $p=0$ as illustrated by the classical action
 \bn  I=\tint(p{\dot q}-\l p)\,dt\;.  \en
The equations of motion read ${\dot q}=\l$, ${\dot p}=0$, and $p=0$. 
On the other hand, one may ask about imposing the constraint $p^3=0$ 
or possibly $p^{1/3}=0$, etc., instead of $p=0$. Let us incorporate 
several such examples by studying the classical action 
 \bn  \tint (p{\dot q}-\l p|p|^\gamma)\,dt\;,\hskip1cm\gamma>-1\;.  \en
Here the equations of motion include ${\dot q}=\l(\gamma+1)\,|p|^\gamma$ 
which, along with the constraint $p|p|^\gamma=0$, may cause some 
difficulty in seeking a classical solution of the equations of motion. 
When $\gamma\neq0$, such constraints are said to be {\it irregular} 
\cite{hen}. It is clear from (\ref{e9}) that irregular constraints 
lead to considerable difficulty in conventional phase-space path integral 
approaches.

Let us examine the question of irregular constraints from the point of 
view of a coherent state, projection operator, phase-space path integral 
quantization. We first observe that the operator $P|P|^\gamma$ is well 
defined by means of its spectral decomposition. Moreover, for any 
$\gamma>-1$, it follows that
  \bn && \hskip-1cm\int e^{-i\xi P|P|^\gamma}{\sin(
\delta^{\gamma+1}\xi)\over\pi\xi}\,d\xi\nonumber\\
  &&=\E(-\delta^{\gamma+1}\le P|P|^\gamma\le\delta^{\gamma+1})\nonumber\\
  &&=\E(-\delta\le P\le\delta)\;.  \en
Thus, from the operator point of view, it is possible to consider the 
constraint operator $P|P|^\gamma$ just as easily as $P$ itself. In 
particular, it follows that
\bn \<p'',q''|\E|p',q'\>
={\cal M}\int\exp[i\tint(p{\dot q}-\l p|p|^\gamma)\,dt]\,\D p\,\D q\,
\D C_\gamma(\l)\;, \en
where we have appended $\gamma$ to the measure for the Lagrange multiplier 
$\l$ to emphasize the dependence of that measure on $\gamma$. The reduction 
of the reproducing kernel proceeds as with the cases discussed earlier, and 
we determine for all $\gamma$ that
\bn \lim_{\delta\ra0}{\sqrt{\pi}\over(2\delta)}\<p'',q''|\E|p',q'\>=
e^{-\half(p''^2+p'^2)}\;,  \en
representative of a one-dimensional Hilbert space. Note that, like the 
classical theory, the ultimate form of the quantum theory is independent 
of $\gamma$. 

It is natural to ask how one is to understand this acceptable behavior 
for the quantum theory for irregular constraints and the difficulties they 
seem to present to the classical theory. Just like the classical and 
quantum Hamiltonians, the connection between the classical and quantum 
constraints is given by 
 \bn \phi(p,q)\equiv\<p,q|\Phi(P,Q)|p,q\>=\<0|\Phi(P+p,Q+q)|0\>\;.   \en
With this rule we typically find that $\phi(p,q)\neq\Phi(p,q)$ due to the 
fact that $\hbar\neq0$, but the difference between these expressions is 
generally qualitatively unimportant. In certain circumstances, however, 
that difference is qualitatively significant even though it is 
quantitatively very small. Since that difference is $O(\hbar)$, let 
us explicitly exhibit the appropriate $\hbar$-dependence hereafter.

First consider the case of $\gamma=2$. In that case
\bn \<p,q|P^3|p,q\>=\<0|(P+p)^3|0\>=p^3+3\<P^2\>p\;,  \en
where we have introduced the shorthand $\<(\cdot)\>\equiv\<0|(\cdot)|0\>$.
Since $\<P^2\>=\hbar/2$ it follows that for the quantum constraint $P^3$, 
the corresponding classical constraint function is given by 
$p^3+(3\hbar/2)p$. For $|p|\gg\sqrt{\hbar}$, this constraint is adequately 
given by $p^3$. However, when $|p|\ll\sqrt{\hbar}$---{\it as must 
eventually be the case in order to actually satisfy the classical 
constraint}---then the functional form of the constraint is effectively 
$(3\hbar/2)p$. In short, if the quantum constraint operator is $P^3$, 
then the classical constraint function is in fact {\it regular} when 
the constraint vanishes. 

A similar analysis holds for a general value of $\gamma$. The classical 
constraint is given by
\bn &&\hskip-1.1cm\phi_\gamma(p)
=({\pi\hbar})^{-1/2}\int(k+p)|k+p|^\gamma e^{-k^2/\hbar}\,dk\nonumber\\
&&=({\pi\hbar})^{-1/2}\int k|k|^\gamma e^{-(k-p)^2/\hbar}\,dk\;. \en
For $|p|\gg\sqrt{\hbar}$  this expression effectively yields 
$\phi_\gamma(p)\simeq p|p|^\gamma$. On the other hand, for $p\approx 0$, 
and more especially for $|p|\ll\sqrt{\hbar}$,  this 
expression shows that the constraint function vanishes {\it linearly}, 
specifically as $\phi_\gamma(p)\simeq \kappa \,p$, where 
  \bn \kappa\equiv 2({\hbar^\gamma/\pi})^{1/2}\int y^2|y|^\gamma e^{-y^2}
\,dy=2({\hbar^\gamma/\pi})^{1/2}\Gamma((\gamma+3)/2)\equiv \hbar^{\gamma/2}
\kappa_o\;. \en
A rough, but qualitatively correct expression for this behavior is given by
\bn  \phi_\gamma(p)\simeq \kappa_o\,p(\hbar+
p^2\kappa_o^{-2/\gamma})^{\gamma/2}\;. \en

Thus, from the present point of view, irregular constraints do not arise 
from consistent quantum constraints; instead, irregular constraints 
arise as limiting expressions of consistent, regular classical constraints 
as $\hbar\ra0$.

\section{OTHER APPLICATIONS OF THE PROJECTION OPERATOR APPROACH}
There have been several cases in which the projection operator has been 
used to study constrained systems.   Shabanov \cite{sha71,sha} as well as 
Govaerts and Klauder \cite{govkla} have applied the projection operator 
formalism to a simple $0+1$ model of a gauge theory. Govaerts \cite {gov1} 
applied the projection operator scheme to study the relativistic 
particle in a reparameterization invariant form. Shabanov and Klauder 
have studied both first-class \cite{sha9} and second-class constraint 
\cite{sha8} situations from the point of view of projection operator 
quantization. In addition, they have discussed in a general way the 
application of projection operator techniques to gauge theory \cite{sha7}. 
Fermion systems have been treated, e.g.,  in \cite{jun}. Shabanov has 
incorporated the projection operator into his Physics Reports \cite{shapr} 
review of gauge theories, and developed an algorithm for how the projection 
operator approach may be incorporated into lattice gauge theory 
calculations. Shabanov has also shown how the projection operator 
approach may be especially useful in ensuring constraints are satisfied 
in an ion-surface interaction \cite{sha98}. In addition, Klauder 
\cite{klagra} has applied the projection operator method in a study of 
quantum gravity. Finally, a U(1) Chern-Simons model has been studied 
and solved with the projection operator method using coherent states in 
\cite{govv}.

Projection operators have also been used previously in the study of 
constrained system quantization. For example, as noted earlier,
some aspects of a coherent state quantization procedure that emphasized 
projection operators for systems with closed first-class constraints have 
been presented by Shabanov \cite{sha71}. In addition, we thank M. Henneaux 
for his thoughtful comments as this approach was being developed, as well 
as for pointing out that projection operators for closed first-class 
constraints also appear in the text of Henneaux and Teitelboim \cite{hen}. 
Please note that this very short list does not pretend to be complete 
regarding prior considerations of projection operator investigations in 
connection with constrained systems.

\section*{Acknowledgements}
The present paper represents a summary of some of the author's principal 
contributions to the projection operator approach for the quantization of 
systems with constraints for the four years from 1996 through 1999. It is 
a pleasure to 
thank my coauthors
J. Govaerts and S. Shabanov who have shared in this general project. In 
addition to these two individuals, thanks are also extended to M. Henneaux 
and B. Whiting for many discussions regarding constraints and their 
quantization.

\end{document}